\documentclass[10pt,journal,compsoc]{IEEEtran}
\usepackage[switch]{lineno}
\usepackage{microtype}                 
\PassOptionsToPackage{warn}{textcomp}  
\usepackage{textcomp}                  
\usepackage{mathptmx}                  
\usepackage{times}                     
\usepackage{tabu}                      
\usepackage{booktabs}                  
\usepackage{soul}
\usepackage{color}
\usepackage{multirow}
\usepackage{hhline}
\usepackage{caption}
\usepackage{booktabs}
\usepackage{subcaption}
\usepackage{hyphenat}
\usepackage{url}
\usepackage{graphicx}

\usepackage[moderate]{savetrees}
\ifCLASSOPTIONcompsoc
  \usepackage[nocompress]{cite}
\else
  \usepackage{cite}
\fi

\ifCLASSINFOpdf
\else
\fi

\begin{document}
%
\title{ViboPneumo: A Vibratory-Pneumatic Finger-Worn Haptic Device for Altering Perceived Texture Roughness in Mixed Reality}
%
%
%
%

\author{Shaoyu Cai, Zhenlin Chen, Haichen Gao, Ya Huang, Qi Zhang, Xinge Yu, and Kening Zhu
\IEEEcompsocitemizethanks{
\IEEEcompsocthanksitem  Shaoyu Cai is with Engineering Design and Innovation Centre, College of Design and Engineering, National University of Singapore.

\IEEEcompsocthanksitem  Haichen Gao, Qi Zhang, and Kening Zhu are with the School of Creative Media, City University of Hong Kong. Kening Zhu is also with CityU Shenzhen Research Institute, Shenzhen, China.%
\IEEEcompsocthanksitem Zhenlin Chen, Ya Huang, and Xinge Yu are with the Department of Biomedical Engineering, City University of Hong Kong.}
\thanks{Corresponding author: Kening Zhu, keninzhu@cityu.edu.hk.}
}

%
%

\markboth{IEEE Transactions on Visualization and Computer Graphics,~Vol.~XX, No.~XX, XXXX~2023}%
{Cai \MakeLowercase{\textit{et al.}}: ViboPneumo}
%



\IEEEtitleabstractindextext{%
\begin{abstract} 
Extensive research has been done in haptic feedback for texture simulation in virtual reality (VR). However, it is challenging to modify the perceived tactile texture of existing physical objects which usually serve as anchors for virtual objects in mixed reality (MR). In this paper, we present ViboPneumo, a finger-worn haptic device that uses vibratory-pneumatic feedback to modulate (i.e., increase and decrease) the perceived roughness of the material surface contacted by the user's fingerpad while supporting the perceived sensation of other haptic properties (e.g., temperature or stickiness) in MR. Our device includes a silicone-based pneumatic actuator that can lift the user's fingerpad on the physical surface to reduce the contact area for roughness decreasing, and an on-finger vibrator for roughness increasing. Our user-perception experimental results showed that the participants could perceive changes in roughness, both increasing and decreasing, compared to the original material surface. We also observed the overlapping roughness ratings among certain haptic stimuli (i.e., vibrotactile and pneumatic) and the originally perceived roughness of some materials without any haptic feedback. This suggests the potential to alter the perceived texture of one type of material to another in terms of roughness (e.g., modifying the perceived texture of ceramics as glass). Lastly, a user study of MR experience showed that ViboPneumo could significantly improve the MR user experience, particularly for visual-haptic matching, compared to the condition of a bare finger. We also demonstrated a few application scenarios for ViboPneumo.
\end{abstract}

}

\maketitle

\IEEEdisplaynontitleabstractindextext

%
\IEEEpeerreviewmaketitle

\IEEEraisesectionheading{\section{Introduction}\label{sec:introduction}}
\begin{figure*}
    \centering
  \includegraphics[width=\linewidth]{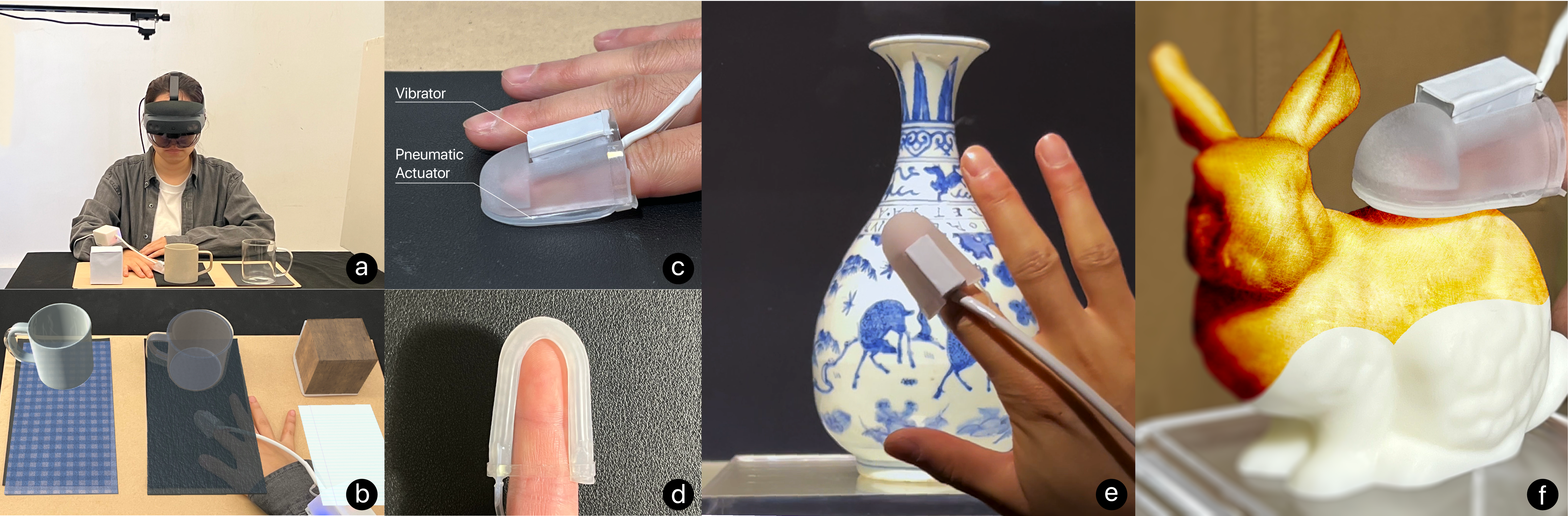}
  \caption{(a) A user wearing ViboPneumo and experiencing the decreased roughness of a virtual leather table mat in MR, 
  (b) A visual scene in MR (HoloLens2) featuring cotton \& leather table mats, glass \& ceramic mugs, a wooden cube, and a piece of paper (c) The detailed view of the user sliding on a texture while wearing ViboPneumo with pneumatic actuation; (d) The bottom view of ViboPneumo with pneumatic actuation; (e-f) Two examples of ViboPneumo applications: altering the perceived roughness of the glass-shelf surface to feel like that of ancient Chinese pottery in the museum, and modifying the perceived roughness of a 3D-printed resin-based object to resemble another material (e.g., copper).}
  \vspace{-0.1in}
  \label{fig:teaser}
\end{figure*}

\IEEEPARstart{T}{he} {texture properties of a material surface, such as roughness, stiffness, and temperature, are crucial information for humans to understand the world \cite{chen2009exploring, djonov2011semiotics}. Through direct touch, we can perceive these properties of a physical object. However, it is still challenging to reproduce these material-related touch sensations in virtual reality (VR) and mixed reality (MR). To this end, researchers have developed various haptic devices or skin-integrated haptic interfaces \cite{yu2019skin} that allow users to interact with virtual objects and feel their tactile properties in VR \cite{bau2010teslatouch, nakagaki2016materiable, bochereau2018perceptual, whitmire2018haptic, degraen2021capturing, cai2020thermairglove, zhu2022tapetouch}. Roughness, in particular, is a crucial factor in determining the tactile perceptual features of a physical surface for texture understanding in the real-world \cite{tiest2010tactual, di2022roughness}. In the existing research on roughness perception in VR, vibrotactile stimuli are commonly deployed on users' hands either through tools (e.g., stylus) \cite{culbertson2014modeling} or directly on fingers \cite{ando2007nail} to simulate the roughness of virtual material surfaces and achieve a considerably realistic haptic experience in VR \cite{preechayasomboon2021haplets, spagnoletti2018rendering}. 
However, it may not be straightforward to apply these VR-focused on-hand vibrotactile stimuli in MR environments where the virtual contents could be overlaid or mounted on the physical world. In such a case, the matching physical object could be used as the interactive haptic proxy to enhance the MR user experience and task performance \cite{kwon2009effects, hochreiter2018cognitive, 10.1145/3313831.3376313}. While the virtual-physical shape matching could be achieved by digital fabrication (e.g., 3D printing), it may be inefficient to fabricate fine surface textures, such as different surface roughness, on the physical proxies for the virtual objects with the same shape but different surfaces. For instance, it might not be cost-effective to 3D-print multiple physical copies of the virtual screws with the same size but different lead distances. Considering the reusability of physical haptic proxies \cite{10.1145/2858036.2858226, 10.1145/3290605.3300923}, 
it would be helpful to use one or fewer physical objects to support the haptic experience of multiple virtual objects with the same physical dimension, but perhaps different textured surfaces. Directly applying vibrotactile stimuli on physical proxies can modify the perceived textures \cite{asano2014toward}, but this approach requires mounting the vibrators on the physical materials. This limitation hinders the scalability and wearability of the haptic interface, as stated in Pacchierotti et al.'s study \cite{pacchierotti2017wearable} on wearable haptic devices. Therefore, it is important to be able to augment or modify the tactile texture sensation of the physical object {through a wearable haptic interface} for MR interaction \cite{bau2012revel}. 

Some researchers have attempted to enrich the touch experience of virtual objects by altering the perceived haptic sensations of physical objects through wearable devices. Vibrotactile feedback is a common approach for increasing the perceived roughness using wearable haptic devices. For example, Asano et al. \cite{asano2013toward} used a ring-based haptic device to add varying frequency vibrations to physical materials, resulting in an increase in perceived roughness. Similarly, electrotactile stimulus has been shown to support perceived roughness increases on the real object's surface for achieving fingerpad-free exploration \cite{yoshimoto2015material}. However, most previous work related to roughness modulation through wearable devices has focused on roughness increasing \cite{ando2007nail, maeda2016wearable, asano2013toward, yoshimoto2015material}. Roughness decreasing is another equally important direction of roughness modulation, but it has received relatively less investigation and may not be trivial \cite{etzi2014textures}. 
Asano et al. \cite{asano2014vibrotactile} deployed a high-frequency on-finger vibrotactile stimulus before the user touched the surface to suppress the skin perceived stimuli of the surfaces, allowing perceived roughness reduction on physical material surfaces. In contrast to the prior work \cite{asano2014vibrotactile}, our work adopts a pneumatic actuator around the fingerpad without blocking the skin, reducing perceived roughness through unobtrusive pneumatic actuation without additional artificial haptic stimuli (e.g., vibrotactile stimuli) before touch interactions. This was inspired by the psychophysical research on human texture perception showing that the reduction of the skin-contact area would lead to the reduction of the applied normal force \cite{van2015review} and further reduce the perceived roughness \cite{lederman1972fingertip}.

In this paper, we present ViboPneumo, an index-finger-worn haptic device that utilizes both vibrotactile and pneumatic feedback to increase or reduce the perceived roughness of physical surfaces while preserving other haptic sensations such as temperature, stickiness, and stiffness, achieving psychological effects of roughness modulation for users. 
As shown in Fig. \ref{Fig.2}, the system includes a linear resonant actuator (LRA), a 3D-printed housing support, a layer of Polydimethylsiloxane (PDMS) film, a layer of Polyethylene terephthalate (PET) film sandwiched between two layers of silicone rubber (EcoFlex 00-30) film, and a pneumatic-vibrotactile control system. The system captures the velocity of the index finger sliding on the physical surface using a top-view camera and generates real-time vibrotactile stimuli with varying frequencies and amplitudes to increase the perceived roughness. The device also includes a hollow pneumatic actuator around the fingerpad to reduce the perceived roughness of the physical surface by lifting the fingertip after inflation, decreasing the contact area between the user's fingerpad skin and the physical surface. With this design, the user's fingerpad is left free to perceive the haptic sensation of the original physical surface, as the user may need to switch interacting with physical and virtual objects in MR. 

We conducted a numerical modeling simulation based on finite element analysis (FEA) to parameterize the inflated displacement/deformation of the proposed pneumatic actuator. Our user-perception {experimental results} showed that ViboPneumo could effectively alter the perceived roughness of various materials, such as glass, ceramics, paper, wood, cotton, and leather. Additionally, we observed that the user-rated roughness ranges might overlap across certain types of materials, indicating the potential to alter the roughness perception from one material to another. For instance, one smooth material (e.g., a glass mug) could be perceived as equally rough as another rougher material (e.g., a ceramic mug). 
Our user studies results of MR experience showed that ViboPneumo could significantly improve users’ MR experience compared to the situation without roughness modulation. According to the users' qualitative feedback, the ViboPneumo system has several potential applications in MR scenarios that require both visual and haptic feedback for exploration (Fig. \ref{fig:teaser}:d-f). For instance, a user could wear the ViboPneumo to explore the surface texture of a museum artifact (e.g., an ancient ceramic vessel) inside a glass shelf, or to select materials and customize products with a free fingerpad for unconstrained texture exploration in MR.


\section{Related Work}
ViboPneumo was inspired by existing research on wearable haptic devices and haptic modulation. 

\subsection{Wearable Haptic Devices}
{Regarding wearable haptic devices for fingers and hands in human-computer interaction (HCI), they can be broadly categorized into two types: glove-type and finger-worn device \cite{pacchierotti2017wearable, maeda2022fingeret}}. Although the glove-type haptic could provide a large area of haptic sensations for users \cite{bouzit2002rutgers, vechev2019tactiles, cai2020thermairglove}, they may suffer from the bulkiness, which may limit their scalability and wearability. In comparison, finger-worn haptic devices offer different types of stimuli (e.g., vibration, force, electrical stimuli, and temperature) with smaller form factors and higher wearability. For example, Tacttoo \cite{withana2018tacttoo} is an electrotactile interface with less than 35$\mu$m thickness, supporting natural tactile cues sensations and ergonomic wearability on the user’s skin. Han et al. developed Hydroring \cite{han2018hydroring}, a finger-worn ring-shaped haptic device that provides pressure, vibration, and temperature sensations on the user's fingertip in MR. Hydroring utilised a thin film of clear Low-Density Polyethylene (LDPE)  as the contact interface and the liquid flow as the actuation, minimizing the interference of fingertip in MR environments interaction. Mazursky et al. presented MagnetIO \cite{mazursky2021magnetio}, a voice-coil actuator on the user's fingernail with interactive soft patches on daily object surfaces, to enhance the touch interaction with every object. Touch\&Fold \cite{teng2021touch} is a nail-mounted haptic device that renders vibration and pressure on the fingertip through a foldable actuator, allowing haptic rendering for virtual objects and interaction for physical objects in MR. More recently, Maeda et al. proposed Fingeret \cite{maeda2022fingeret}, a fingerpad-free haptic device that enables vibrotactile feedback for seamless interactions through two actuators on the fingernail and finger side in MR. These finger-mounted haptic devices can provide various haptic feedback, potentially supporting the rendering of different haptic properties of material surfaces, such as roughness, slipperiness, stiffness, and temperature.

Researchers have developed various wearable haptic technologies to render the roughness or friction of material surfaces. Vibrotactile display is a low-cost approach to present the rich tangible vibrations to simulate the surface textures during the interaction with the physical objects \cite{choi2012vibrotactile, strohmeier2017generating}. 
Some researchers explored tool-mediated texture exploration for roughness perception, providing different roughness experiences via frequency and amplitude adjustments. For example, Kyung and Lee presented Ubi-Pen \cite{kyung2008ubi}, a pen-based haptic interface providing vibration and pressure to render the roughness of textured surfaces from image data. Such an approach extracted the grey values of an image and mapped them to the intensity of vibrotactile stimuli on each pin of the tactile display. To further provide realistic texture sensations, Romano et al. \cite{romano2011creating} adopted a data-driven approach, which generates texture roughness from acceleration data captured from physical material surfaces through a vibratory stylus device. In addition to tool-based texture exploration, some existing works have simulated roughness on bare skin. Schorr and Okamura \cite{schorr2017fingertip} proposed a pair of finger-mounted haptic devices that stretch the skin of fingertips for stiffness and friction rendering. Chen et al. \cite{chen2019fw} presented FW-Touch, a wearable haptic device that displays hardness, friction, and roughness on a virtual surface through force and vibrotactile feedback on the finger to simulate the surface material properties on a touch screen. FinGAR \cite{yem2017wearable} used vibration and electrical stimulation to produce skin deformation and vibration for roughness and hardness simulation. 

Despite the extensive research efforts on wearable haptic devices for roughness simulation in VR/MR, most of them (except Touch\&Fold \cite{teng2021touch} and Fingeret \cite{maeda2022fingeret}) cover the user's fingerpad, preventing direct interaction with real-world objects. This may not be suitable for the MR scenario where users may interact with the virtual and the physical objects alternatively. Touch\&Fold \cite{teng2021touch} and Fingeret \cite{maeda2022fingeret} are fingerpad-free devices, but they did not attempt to support roughness modulation on physical surfaces, and only provided haptic feedback for either virtual or physical objects binarily in MR. In our research, ViboPneumo can alter the user-perceived roughness for real-world objects, and provide the corresponding haptic textured sensations for the virtual objects overlaid on the physical objects in MR.

\begin{figure}[ht] 
\centering 
\includegraphics[width=\columnwidth]{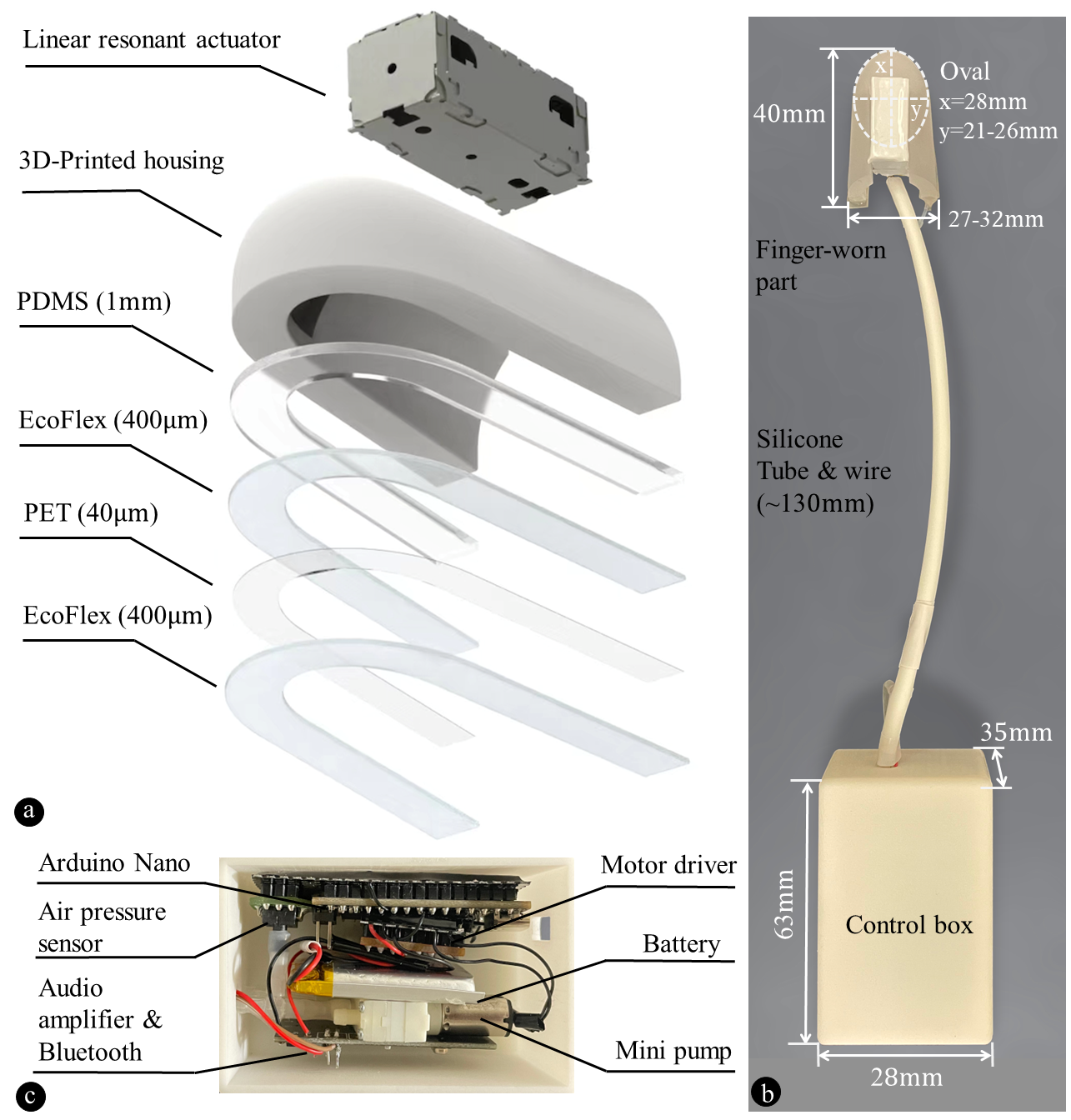} 
\caption{(a) {The structure of ViboPneumo with dimensional information}: A linear resonant actuator and a 3D-printed semi-transparent housing with a layer of PDMS film with 1mm thickness, two layers of silicone rubber that were 400 $\mu$m thick, and a layer of PET film (40$\mu$m thickness) in the middle; (b) The appearance of the wearable device; (c) All the electronic components were assembled into a control box as the pneumatic-vibrotactile control system.} 
\vspace{-0.2in}
\label{Fig.2}  
\end{figure}

\subsection{Haptic Modulation}
Haptic modulation refers to presenting synthetic signals generated from haptic actuators to alter the physical haptic properties of real materials \cite{jeon2009haptic}. As an example, Jeon and Choi \cite{jeon2009haptic} proposed Haptic Augmented Reality (HAR), altering the perceived stiffness of a real object by combining the PHANToM-generated force stimuli with the original stiff force from the real object. Similarly, Hachisu et al. \cite{hachisu2012augmentation} proposed a HAR system that altered the perceived stiffness of a real object through vibration to support users' material identification. Later, Tao et al. \cite{tao2021altering} developed a wearable haptic device that supported the stiffness reduction of a real rigid object by restricting the lateral deformation of the fingerpad through a hollow frame. 
These HAR systems could not only modulate the stiffness of real objects but also preserve the texture perception for the objects, which is important for surface understanding and object identification with touch \cite{chen2009exploring, djonov2011semiotics}.  

One major research direction of haptic modulation is to alter the perceived roughness of real-world materials to create the haptic sensations of new materials. Several existing research in psychology and cognitive science have revealed that roughness is one of the most important features for haptic exploration and discrimination on material-textured surfaces \cite{holliins1993perceptual, hollins2000individual, tiest2006analysis, tiest2010tactual}. 
Asano et al. \cite{asano2012vibrotactile, asano2014toward} directly adopted vibrotactile stimuli to augment the perceived roughness of the real-world textured surfaces (e.g., cloth, paper and wood), to create the haptic sensations of different virtual materials. 
Then, the same group investigated two different types of vibrotactile stimuli effects for texture modulation: finger-ring-based vibrating (i.e., the vibrator was worn on a finger) and material-based vibrating (i.e., the material was mounted on a vibrator) \cite{asano2013toward}. The results showed that the finger-ring-based vibration method was preferred for AR application, and was applied for their later work in roughness modulation as a wearable texture display \cite{asano2014vibrotactile}. Yoshimoto et al. \cite{yoshimoto2015material} adopted the electrotactile stimuli on the user's finger to increase the roughness of real materials. 
These aforementioned works demonstrate that vibrotactile or electrotactile stimuli can be used to modulate or modify the user-perceived roughness of real-world textured surfaces. However, their studies mainly focused on roughness increasing roughness rather than decreasing, and this may limit the scalability of the haptic modulation system. 
In terms of roughness reduction, Asano's work \cite{asano2014vibrotactile} revealed that introducing high-frequency vibratory stimuli on the finger skin before contacting the object surface could decrease the perceived texture roughness. This is because the high-frequency on-finger vibration may reduce the sensitivity of the tactile receptors on the skin. However, this approach introduced additional haptic stimuli before the user's actual surface interaction, potentially leading to an obtrusive interaction experience. Besides the motor-based vibrotactile stimuli, the ultrasonic vibration could be another solution for roughness modulation, especially for reducing the touch sensation of bumpy textures on a physical surface  \cite{ochiai2014diminished}. This is based on the effect of squeezing air film between the finger skin and the surface. 
However, this approach mounted the ultrasonic transducer on the material sample, limiting the system scalability and wearability which is important for haptic interfaces \cite{pacchierotti2017wearable}. 

In our work, without instrumenting the real-world objects, ViboPneumbo is a finger-worn haptic device that combines both vibrotactile stimuli and pneumatic actuation to increase and decrease the roughness of physical material surfaces, respectively. With ViboPneumo, we aim to improve the visual-haptic matching experience in MR in an unobtrusive and fingerpad-free manner. Different from the previous work on reducing the perceived roughness, we do not apply any artificial haptic stimuli on the user's hand \cite{asano2014vibrotactile} or the to-be-touched surface \cite{ochiai2014diminished} before the touch action. 

\section{System Design and Implementation}
ViboPneumo is a lightweight and fingerpad-free wearable haptic device that generates both pneumatic actuation and vibrotactile feedback for modulating the perceived roughness of textured surfaces. The combination of these two types of haptic feedback achieves the effects of increasing and decreasing the perceived roughness, respectively. Previous work has shown that vibrotactile feedback is a low-cost and wearable approach commonly adopted for roughness increasing \cite{asano2012vibrotactile, asano2013toward, ando2007nail}. However, achieving the effect of roughness decreasing with current hand/finger-worn devices may not be straightforward. {Research related to human perception and cognition has shown that the contact area between the fingerpad and a textured surface is positively proportional to the applied normal force \cite{van2015review}, which in turn is positively correlated with roughness \cite{lederman1972fingertip}. Therefore, we hypothesized that it might be possible to decrease perceived roughness by reducing the contact area. To test this hypothesis, we decided to design a finger-worn device that could reduce the contact area between the fingerpad and physical surfaces.}

{One extreme approach to reducing the contact area would be to fully block the fingerpad's contact with the physical surface using a smooth cover, such as acrylic or rubber. However, deploying haptic actuators between humans' fingerpads and contact objects to fully cover the fingerpad may limit users' direct perception of haptic attributes of physical objects, such as stickiness or temperature. This is not appropriate for mixed-reality interactions, which require seamless interactions with both physical and virtual objects \cite{maeda2022fingeret}. While it is possible to develop thin wearable haptic actuators that cover fingerpads and allow users to feel through the physical properties of real objects \cite{withana2018tacttoo}, users may still feel these thin haptic actuators, which could limit their direct perception of textured surfaces and lower their overall user experience \cite{nittala2019like}. Furthermore, haptic actuators that fully block users' fingerpads would limit haptic augmentation on real objects, which can enhance the daily haptic experience of real objects, such as increasing the perceived roughness of a fabric \cite{asano2012vibrotactile, maeda2016wearable}.}

{Given that humans' palms and fingerpads are crucial for dexterous manipulations and interactions with physical objects through touch \cite{shimoga1993survey}, it is not trivial to keep users' fingerpads free to support the direct perception of textured surfaces with preserving manual dexterity in VR/MR. Inspired by recent work on rendering electrotactile feedback without obstructing the palmar side of the hand in VR \cite{tanaka2023full}, we decided to deploy an around-fingerpad pneumatic actuator and drive it to lift the fingertip to reduce the contact area with physical objects, thus decreasing perceived roughness. We also incorporated a vibrator on the top of the fingertip to increase perceived roughness. This design enables fingerpad-free interaction with direct perception of the texture of physical objects while providing roughness modulation in MR and preserving manual dexterity.}

\begin{figure}[ht] 
\centering 
\includegraphics[width=\columnwidth]{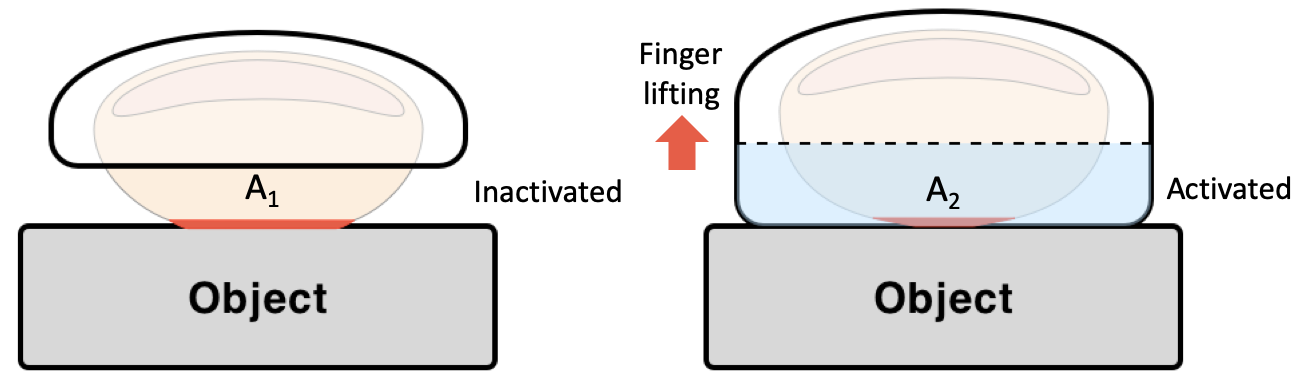} 
\caption{The principle of reducing contact area through pneumatic actuation. When the pneumatic actuator is activated (right), it lifts the fingertip and causes a decrease in the contact area ($A_2$) compared to the contact area ($A_1$) when the actuator is inactive (left).} 
\vspace{-0.15in}
\label{Fig.3}  
\end{figure}

\subsection{Pneumatic-Actuator Design and Fabrication}


To reduce the contact area of the fingerpad on the textured surface, we designed a U-shape pneumatic actuator around the edge of the fingerpad but does not fully cover it, allowing for fingerpad-free interaction. As shown in Fig. \ref{Fig.3}, the pneumatic actuator can be inflated to lift the fingerpad. We used silicone as the medium in contact with the physical textured surface for the pneumatic actuator. Additionally, we covered an additional layer of Polydimethylsiloxane (PDMS), which is a silicone elastomer, on the pneumatic actuator to make its upper layer expandable and for easy assembly with a 3D-printed housing support through silicone adhesive.

\begin{figure*}[htbp] 
\centering 
\includegraphics[width=\textwidth]{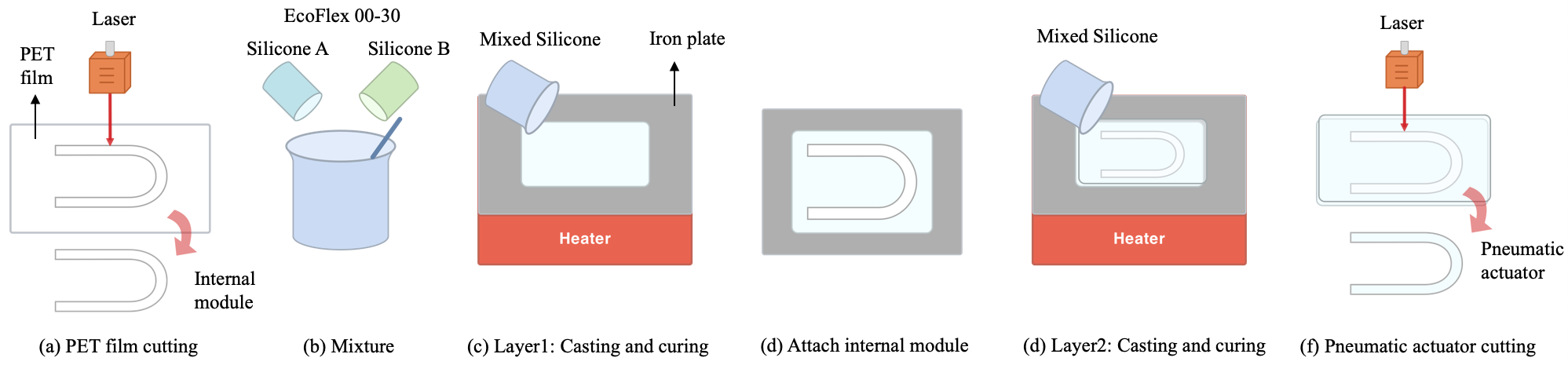} 
\caption{The process of pneumatic-actuator fabrication.}
\label{Fig.4} 
\vspace{-0.15in}
\end{figure*}

{Fig. \ref{Fig.4} illustrates the process of pneumatic-actuator fabrication.} To fabricate the pneumatic actuator, we first laser-cut a 40 $\mu$m thickness layer of transparent thermoplastic polyester (PET) film to obtain a U-shape internal module. Next, we mixed the two parts of liquid silicone rubber (EcoFlex 00-30 A and B) with a 1:1 ratio and stirred for about a minute. We then poured the silicone mixture onto an iron plate and used a film applicator to spread the mixture to a thickness of 400 $\mu$m. The silicone was then heated for about 20 minutes to cure. After the silicone film was cured, we attached the laser-cut U-shaped internal module to the first layer of the dry silicone film. We then repeated the casting and curing procedure to create another layer of silicone film on top. This process created a non-cured chamber between the two layers of silicone film that could be inflated with different volumes of air pumping. For details on the fabrication process, please refer to our supplementary video (01:30 - 02:00).

\subsection{Finger-Wearable System Design}
Fig. \ref{Fig.2}a shows the structure of ViboPneumo. According to our pilot study, we adopted the aforementioned pneumatic actuator which was attached to a 3D-printed semi-transparent housing structure made with Thermoplastic Polyurethane (TPU). Additionally, a linear resonant actuator (LRA) (20 $\times$ 12 mm) was fixed on the top of the housing structure, for rendering the increased roughness on the textured surface (Fig. \ref{fig:teaser}c).
We fabricated different sizes of the U-shaped pneumatic actuator and the 3D-printed housing structure to accommodate different finger sizes. The width of the pneumatic tube is 4 mm, and we determined the width of the inner hollow ellipse for the pneumatic actuator to be 13, 14, 15, 16, 17, and 18 mm for different finger sizes. The Polytetrafluoroethylene (PTFE) tube was embedded between two layers of silicone rubber film and connected to a control box (Fig. \ref{Fig.2}b) including a pneumatic-pump control system for air pressure control, and the LRA was connected to the vibratory-signals generation system for frequency controlling and vibrotactile rendering (Fig. \ref{Fig.2}c).

\subsubsection{Vibrotactile Signal Generation}

Adopting the similar methods proposed by Asano et al. \cite{asano2013toward, asano2012vibrotactile}, we installed a linear resonant actuator on the fingernail to generate vibrotactile feedback for perceived roughness increases. 
We employed the concept of virtual wavy surface which has been applied for virtual texture rendering \cite{okamura1998vibration, bau2010teslatouch} and roughness modulation \cite{asano2012vibrotactile, asano2014vibrotactile, ujitoko2019modulating}. Specifically, the system utilized a linear resonant actuator to generate sinusoidal vibratory stimuli {that were oriented parallel to the index finger for better perception \cite{brisben1999detection} based on the estimated finger velocity during the finger scanning on the textured surface. It is worth noting that the previous research \cite{landin2010dimensional} has shown that humans have difficulty distinguishing the direction of high-frequency vibrations.} The value of the driven voltage \textit{Y(t)} was determined by the following equation:

\begin{equation}
\mathop{\textit{Y(t)} = A \sin (2 \pi \frac{\textit{v(t)}}{\lambda} + \varphi})
\label{eq1}
\end{equation}

where \textit{Y(t)}, $A$, \textit{v(t)}, and $\lambda$ were the driven voltages, the vibratory amplitude, the velocity of the index finger movements on the material surface, and the wavelength of the rendered virtual surface, respectively. In our case, $\lambda$ was set constant as 1.0 mm, {and phase value $\varphi$ was set as 0,} followed in the previous work \cite{asano2014vibrotactile}.

\subsubsection{Pneumatic-Vibrotactile Control System}
Fig. \ref{fig:system} depicts the flow diagram of our pneumatic-vibrotactile control system. We used a miniature vacuum pump (SC3101PW, DC 3.0 V, 65 mA, 33.5 mm $\times$ 10mm $\times$ 4mm) 
as our air-pumping source. The control system also included an air-pressure sensor (XGZP6847A, CFSensor) with a sampling frequency of 20 Hz for the closed-loop pneumatic control. The air pump was controlled by an external motor-driver circuit (TB6612FNG) with a 3.7 V, 200 mAh lithium polymer battery. For the vibrotactile control, a power amplifier (5W, M38) communicated with the computer through Bluetooth and drove the linear resonant actuator. 

The MR application detected the object-touching action and triggered the roughness-modulation process. Specifically, for roughness decreases, taking the referenced air pressure as the set point and the measured air pressure reading of the air pressure sensor as the feedback, the micro-controller (Arduino Nano) controlled the pneumatic pumping system to achieve and maintain the target air pressure in the pneumatic actuator. The top-view camera could capture the velocity of the index finger, and the computer generates the vibration signals and transmits the signals to the linear resonant actuator through the amplifier for vibrotactile rendering.

\begin{figure}[ht]
 \centering 
 \includegraphics[width=\linewidth]{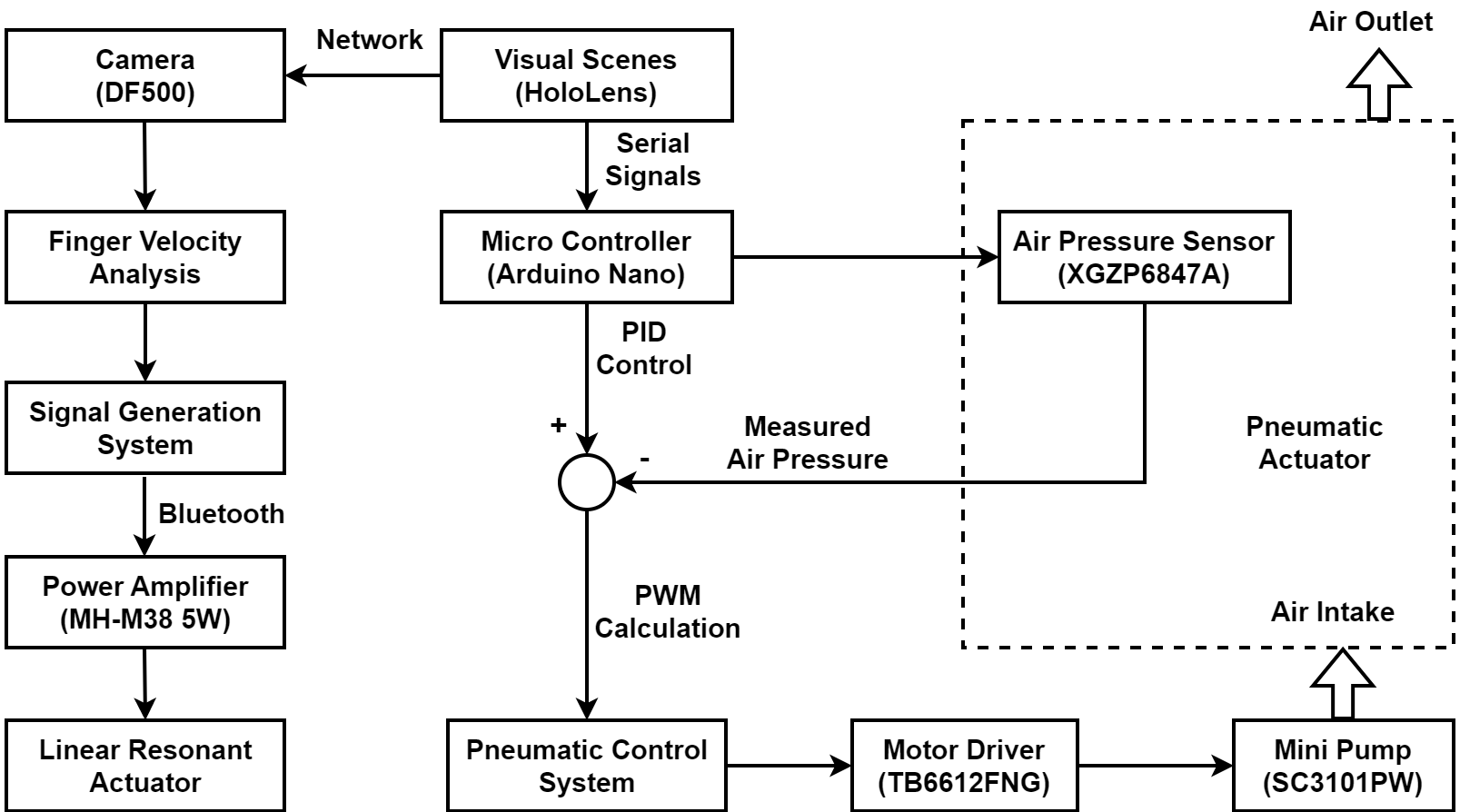}
 \caption{The pneumatic-vibrotactile control system.}
 \vspace{-0.15in}
 \label{fig:system}
 \end{figure}

\subsection{Software Implementation}

We developed the MR scenarios using  Unity3D 2019.4.39f with C\#, and deployed them on a HoloLens 2 device. We used OpenCV \cite{bradski2000opencv} to detect the user's hand movements and calculate the velocity of the index finger, and the PyAudio library\footnote{https://people.csail.mit.edu/hubert/pyaudio/} for the vibrotactile signal generation.
Specifically, following the similar processing of Asano's work \cite{asano2014vibrotactile}, we adopted a hand-tracking algorithm based on OpenCV to extract the positions of hand landmarks with a sampling rate of 30 Hz and estimated the current velocity of the index fingertip under a 50 cm height distance between the top-view camera and the physical surface at 3000 Hz. We then mapped the current velocity values into the wave of driven voltages through a sine wave function and played the vibrotactile signals as an audio sequence to drive the linear resonant actuator with varying frequencies at a 1000Hz frame rate. When the computer received the control signals from HoloLens 2 (i.e., the trigger of hand-object contact), the software controlled the linear resonant actuator and the pneumatic system to increase and decrease the perceived roughness of physical materials respectively. 

\section{Technical Evaluation}

As we deployed the commercial vibrotactile product with standard control circuitry, we mainly evaluated our customized pneumatic system. Specifically, we examined the performance of the closed-loop air-pressure control, the lifted distances of the fingerpad under different inflation volumes, the reduced finger-surface contact areas, and the other general system characteristics (i.e., latency, response time, power consumption).

\subsection{Air Pressure Tracking}

We first evaluated ViboPneumo's performance in generating different levels of air pressure in the U-shape pneumatic tube. 
Before formal testing, we noticed uneven deformation on the pneumatic tube when the air pressure was over 12 kPa, which may be due to the minor variance of the thicknesses across different parts of the tube caused by the uneven heat distribution during the curing process.
To this end, we measured the step responses of target air pressure from 0 to 12 kPa with 1 kPa as the interval, a total of 12 controlled pneumatic signals. We implemented the PID control algorithm with a sample period of 0.05s for real-time inflating and deflating. To demonstrate the system's air-tracking performance, Fig. \ref{fig:track} illustrates the step responses of our system to three air pressure (6, 8, and 10 kPa). We also calculated the mean absolute errors (MAE) and the maximum measured errors (MME) of the air pressure-changing proportional stage (i.e. before reaching the target air pressure) and the stable stage (i.e. 5s for maintaining the air pressure), respectively. The results showed that the average MAE was 0.679 kPa and MME was 1.233 kPa in the proportional stage, 0.386 kPa for MAE and 0.990 kPa for MME in the stable stage. Table 1 in Supplementary Material of ViboPneumo shows all MAEs and MMEs of each reference air pressure (0-12 kPa).

\begin{figure}[h]
 \centering 
 \includegraphics[width=\columnwidth]{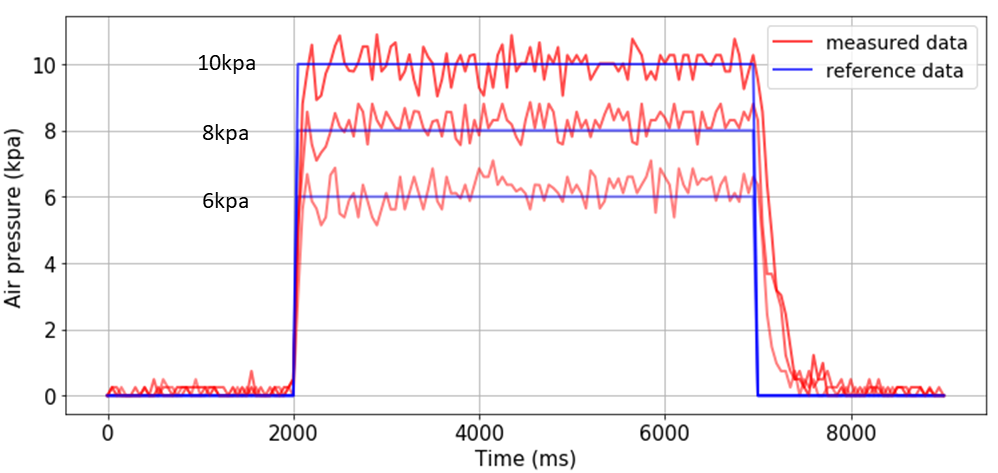}
 \caption{The air-pressure tracking results for the PID control system.}
 \vspace{-0.15in}
 \label{fig:track}
 \end{figure}

\begin{table}[ht]
\caption{The MAE and MME values of different reference air pressure values.}
\resizebox{\columnwidth}{!}{\begin{tabular}{|c|cc|cc|}
\hline
\multirow{2}{*}{\begin{tabular}[c]{@{}c@{}}Reference air \\ pressure (kPa)\end{tabular}} & \multicolumn{2}{c|}{Proportional stage}    & \multicolumn{2}{c|}{Stable stage}          \\ \cline{2-5} 
                                                                                         & \multicolumn{1}{c|}{MAE (kPa)} & MME (kPa) & \multicolumn{1}{c|}{MAE (kPa)} & MME (kPa) \\ \hline
1                                                                                      & \multicolumn{1}{c|}{ 0.48}        &   0.91         & \multicolumn{1}{c|}{0.48 }            &    0.88    \\ \hline
2                                                                                        & \multicolumn{1}{c|}{0.61}         &      1.18     & \multicolumn{1}{c|}{0.39}          &      0.94     \\ \hline
3                                                                                        & \multicolumn{1}{c|}{0.45}          &      0.97     & \multicolumn{1}{c|}{0.38}          &     0.91      \\ \hline
4                                                                                        & \multicolumn{1}{c|}{0.45}          &    1.10       & \multicolumn{1}{c|}{0.40}          &      1.11     \\ \hline
5                                                                                        & \multicolumn{1}{c|}{0.23}          &    0.74       & \multicolumn{1}{c|}{0.34}          &       1.12    \\ \hline
6                                                                                        & \multicolumn{1}{c|}{0.51}          &     0.86      & \multicolumn{1}{c|}{0.41}          &      1.09     \\ \hline
7                                                                                        & \multicolumn{1}{c|}{1.04}          &    1.31       & \multicolumn{1}{c|}{0.31}          &        0.89   \\ \hline
8                                                                                        & \multicolumn{1}{c|}{0.70}          &    1.14       & \multicolumn{1}{c|}{0.38}          &      0.85     \\ \hline
9                                                                                        & \multicolumn{1}{c|}{0.98}          &    1.12       & \multicolumn{1}{c|}{0.43}          &       1.08    \\ \hline
10                                                                                       & \multicolumn{1}{c|}{0.72}          &      1.20     & \multicolumn{1}{c|}{0.35}          &      0.97     \\ \hline
11                                                                                       & \multicolumn{1}{c|}{1.00}          &      2.2     & \multicolumn{1}{c|}{0.40}          &       1.12    \\ \hline
12                                                                                       & \multicolumn{1}{c|}{0.99}          &     2.07      & \multicolumn{1}{c|}{0.37}          &       0.92    \\ \hline
Average                                                                                  & \multicolumn{1}{c|}{0.679}          &    1.233       & \multicolumn{1}{c|}{0.386 }         &      0.990     \\ \hline
\end{tabular}}
\label{tab:maemme}
\end{table}

\subsection{Pneumatic Deformation}

To evaluate the deformation of the U-shape pneumatic tube, we built a 3D finite element model (FEM) in COMSOL Multiphysics 5.5 (COMSOL, Sweden) to simulate different levels of shape-changing under air pressures. We restructured and simplified the 3D model of the pneumatic tube as a U-shape plate consisting of two layers of stretchable silicone film with a thickness of 400 $\mu$m (Ecoflex-30, Smooth‑On, USA) and applied a set of pressure loading in the tube from 0 to 12 kPa.  We adjusted the elastic parameters of the silicone material (e.g., elastic modulus) according to the deformation response from our measurement of the expanded distance, resulting in linear relations between the height distance and the air pressure, as shown in Fig. \ref{fig:simulationresult}a. Fig. \ref{fig:simulationresult} also shows our experimental measurement of the lifted distances in the real pneumatic tube under different air pressure, indicating that our air-pressure control system could closely follow the simulation. Moreover, Fig. \ref{fig:simulationresult}b shows that the maximum deformation of the pneumatic tube was evenly distributed in the middle of the pneumatic channel, implying that fingers could be lifted smoothly. Fig. \ref{fig:simulationresult}c shows that the distribution of stress continued from the middle to the edge of the pneumatic tube during deformation. {The deformation and the stress directions are perpendicular and tangent to the U-shape surface, respectively.}

 \begin{figure}[htbp]
 \centering 
 \includegraphics[width=\linewidth]{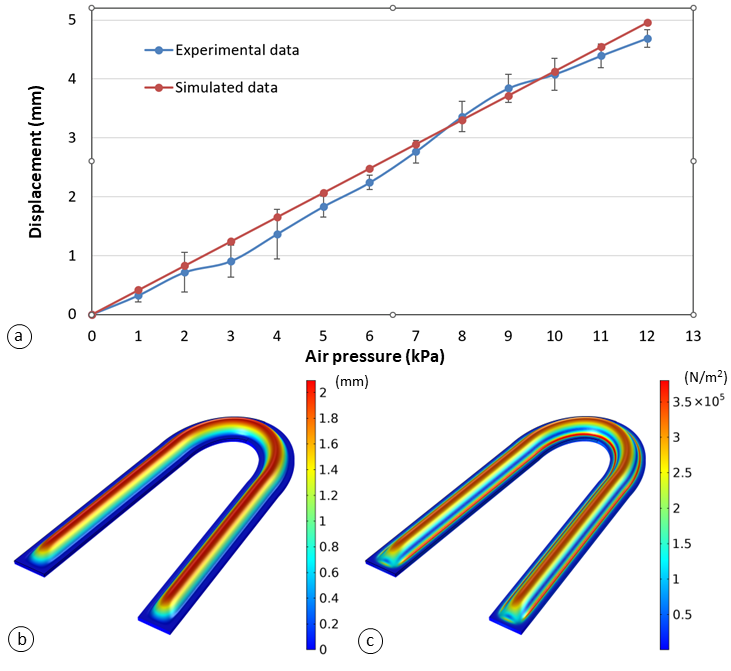}
 \caption{The FEM analysis of the U-shape pneumatic tube. (a) The simulated and experimental data; (b) The static deformation and, (c) The stress distribution (both under 6 kPa).}
 \label{fig:simulationresult}
 \end{figure}
 

\subsection{Reduced Finger-Surface Contact Area}
By lifting the fingerpad, our device could potentially reduce the contact area between the user's fingerpad and the physical textured surface. To this end, we recruited three participants to wear the ViboPneumo device on their index fingers and press their fingertips on a paper surface with an applied force of roughly 0.75 N, 1.0 N, and 1.5 N. We collected their fingerprints with ink on an ergometer (DS2-5N, PUYAN) and tested three different air-pumping pressures: 6 kPa, 8 kPa, and 10 kPa, to compare their finger-surface contact areas with that of no air-pumping stimuli. Fig. \ref{area} shows the images of one participant's fingerprints with the edge lines. We also calculated the pixel numbers of each image of fingerprints and calculated the ratio of contact areas with/without pneumatic stimuli. {The results indicate that the ratio of reduced contact area increased as the air pressure increased and the normal force decreased. Specifically, the average values of the ratio of reduced contact area were as follows: 8.1\% (SD = 2.15\%) for 6 kPa, 11.9\% (SD = 3.18\%) for 8 kPa, and 20.8\% (SD = 2.62\%) for 10 kPa at a normal force of 1.5 N; 8.6\% (SD = 4.46\%) for 6 kPa, 15.2\% (SD = 3.20\%) for 8 kPa, and 28.8\% (SD = 6.12\%) for 10 kPa at normal force of 1.0 N; and 12.8\% (SD = 3.67\%) for 6 kPa, 25.5\% (SD = 2.55\%) for 8 kPa, and 39.6\% (SD = 3.79\%) for 10 kPa in the 0.75 N normal force condition.}

\begin{figure}[ht]
 \centering 
 \includegraphics[width=\columnwidth]{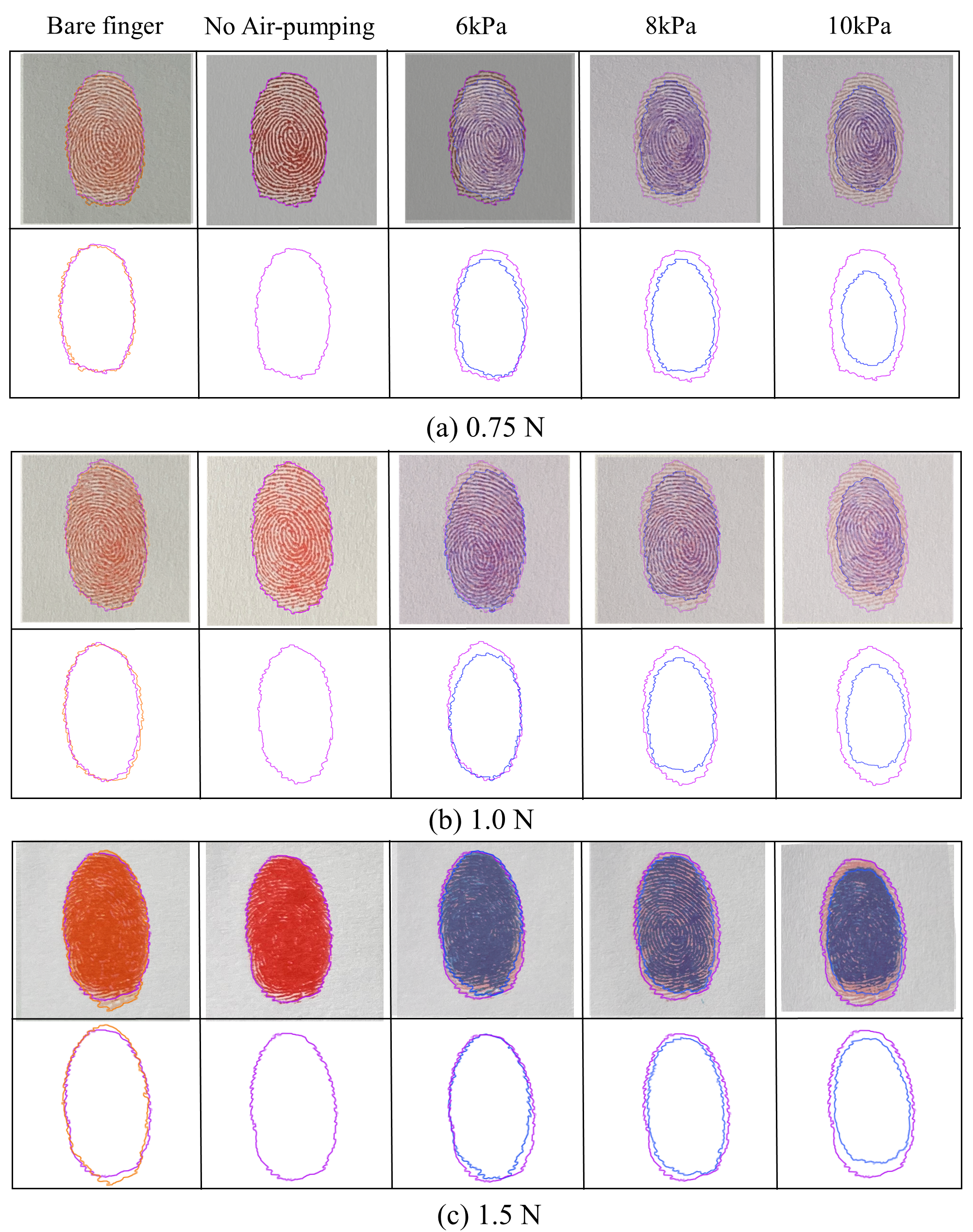}
 \caption{One participant's fingerprints under different {applied normal forces: (a) 0.75 N, (b) 1.0 N, and (c) 1.5 N}. For each sub-figure, the first row includes the fingerprints printed on the paper (orange for the bare finger, red for no air-pumping and blue for air-pumping stimuli). The second row illustrates the edges of fingerprint images extracted by PhotoShop (The orange line for the bare finger, the purple line for no air-pumping and the blue lines for different air-pumping stimuli). }
 \label{area}
 \vspace{-0.1in}
 \end{figure}
 
\subsection{Latency, Noise, and Power Consumption}
We also measured the response time for our system. 
The results revealed an average activation time of 145.83 ms (SD = 39.65) and 329.17 ms (SD = 94.05) for deactivation, suggesting a real-time performance as shown in Fig. \ref{fig:track}. The latency of vibrotactile feedback is approximately 53.53 ms (SD = 8.32) under the refreshing rate of 30Hz refresh rate for the process of hand-motion tracking and velocity estimation with a Dell i5 CPU. 
To assess the noise level of ViboPneumo, we placed a sound-level meter at a distance of about 40 cm from the ViboPneumo, simulating the approximated distance between the user’s wrist and ear in usage. The measured noise levels were about 51.0 dB for 6 kPa, 56.9 dB for 8 kPa, and 59.0 dB for 10 kPa control signals, with the ambient noise level at approximately 38.9 dB. We powered our device using a 200 mAh, 3.7V LiPo battery. The highest current is about 60 mA (0.2W) for 10 kPa pneumatic actuation and the highest power consumption for vibratory is about 2 W under 6.2 m/$s^2$ acceleration signals in 250 Hz. Therefore, the maximum total power consumption of the current system is about 2.2 W, which can support approximately 19 minutes of continuous tactile feedback. 

\section{User-perception Experiment 1: Modulating Surface Haptic Properties by ViboPneumo}

After technically testing ViboPneumo, we conducted a user-perception experiment to investigate how our device can alter users' perception of surface haptic sensations on different materials. We considered all five psychophysical dimensions of haptic interaction, namely roughness, flatness, temperature, stickiness, and stiffness \cite{okamoto2012psychophysical}. 
{Noted that the perceived roughness of the material surface is mediated by the vibrational cues (spatial periods smaller than 200$\mu$m) for fine roughness and spatial cues for flatness or macro roughness (spatial periods exceeding 200$\mu$m) \cite{hollins2000evidence}. In terms of subjective representation, the flatness was perceived and represented by the ``uneven'' label, and the fine roughness was mainly described as ``rough'' \cite{okamoto2012psychophysical}.} We hypothesized that the perceived roughness or flatness of the textured surfaces might be increased through ViboPneumo's vibrotactile feedback \cite{asano2012vibrotactile}, and {decreased due to the reduction of finger-surface contact area \cite{van2015review} and applied normal force \cite{lederman1972fingertip} on the surface, through the pneumatic actuation}. Furthermore, we were interested in studying how ViboPneumo may affect the users' perception of other psychophysical haptic dimensions (i.e., temperature, stickiness, and stiffness). {The experiment protocol was approved by the ethical board of the university.}

\subsection{Apparatus and Stimuli}

Fig. \ref{fig:material1}a illustrates the setup of the study environment, including the ViboPneumo system, a Microsoft Surface Pro for showing the graphical user interface (Fig. \ref{fig:material1}b) and recording the participants' rating responses. The participant wore ViboPneumo on his/her dominant hand which was placed behind a large cardboard to avoid visual bias. An arm support was fixed on the table, to reduce possible fatigue during the experiment. The participant also wore a pair of earmuffs to avoid auditory bias. A top-view camera (5 MP, 30 fps, JERRY, CHINA) was installed at the height of 50 cm and captured the index-finger motion for velocity calculation. 

We selected seven different materials (glass plate, ceramics plate, paper, plywood, balsa wood, taurillon leather, and cotton, as shown in Fig. \ref{fig:material1}c) for this user-perception experiment, to study how ViboPneumo may alter users' haptic perception on these materials. {We selected four materials (leather, paper, wood, and cotton) based on previous work related to roughness/texture modulation \cite{asano2013toward, yoshimoto2015material}. To balance the range of roughness, we added two smoother materials: glass and ceramics and plywood material with medium roughness.} All the material samples were cut into the size of 10 cm $\times$ 10 cm with a thickness of 2 mm, and placed on a 10 cm height acrylic shelf.

\begin{figure}[htbp]
 \centering 
 \includegraphics[width=\linewidth]{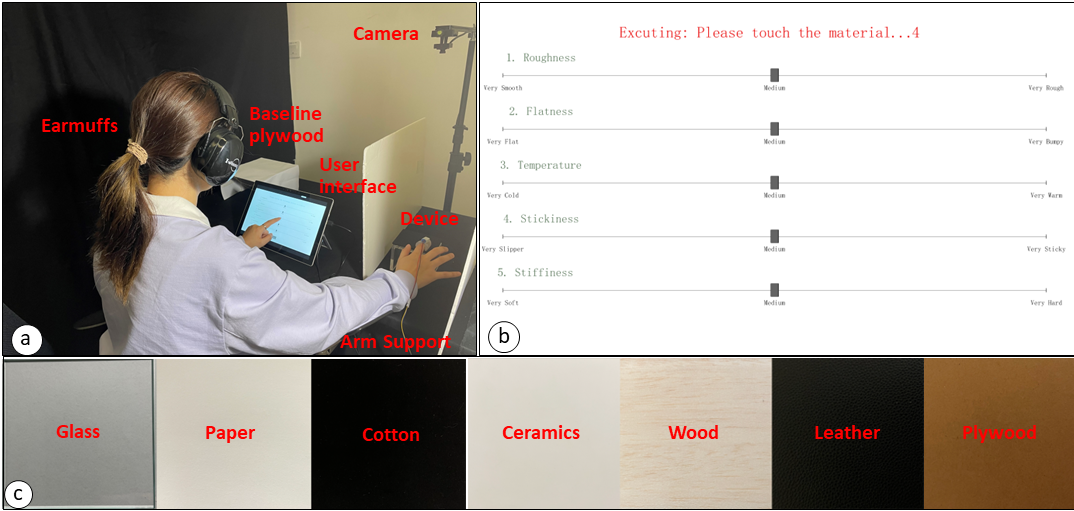}
 \caption{(a) Setup of the study environment, (b) the user interface, (c) selected six testing materials and baseline material.}
 \vspace{-0.1in}
 \label{fig:material1}
 \end{figure}

The set of haptic stimuli for ViboPneumo includes three types of vibrotactile stimuli, three types of pneumatic stimuli, and one condition with no stimuli, for a total of seven types of stimuli. We followed the similar amplitude settings of vibrotactile stimuli for roughness increasing in Asano's work \cite{asano2014vibrotactile}. Specifically, the three amplitude levels of vibrotactile stimuli corresponded to 3.7, 4.9, and 6.2 m/${s}^2$ acceleration at 250 Hz, denoted as the stimuli A1, A2 and A3, respectively. Higher amplitude tended to yield rougher perception when the user scanned his/her finger on a textured surface \cite{strohmeier2017generating}. Besides, we selected three different levels of air pressure for pneumatic actuation: 6 kPa, 8 kPa, and 10 kPa, where the corresponding fingerpad-lifting distances were averagely 2.24 mm (SD = 0.121), 3.36 mm (SD = 0.258) and 4.07 mm (SD = 0.274) as shown in our technical experiments. These stimuli were denoted as stimuli B1, B2, and B3, respectively. 

Before the experiment started, we asked all the participants to rank the perceived roughness levels of these seven materials (i.e., 1 - smoothest and 7 - roughest) with only their bare fingers touching the surfaces and without wearing ViboPneumo. The results showed that glass was consistently ranked as the smoothest, and over 80\% of the participants ranked leather as the roughest, with plywood always ranked as the fourth smooth/rough material out of seven. Hence, we chose plywood as our baseline material, and the other six types of materials as the testing materials. That is, the participant was instructed to rate their perceived roughness compared to their non-instrumented perception of plywood roughness, which indicated the center position of the rating slider in the graphical user interface on the Surface Pro device.

\subsection{Experimental Design}
We employed a within-subject factorial design for this experiment, with two independent variables: the type of haptic stimuli (i.e., the six ViboPneumo signals and the no-stimulus situation) and the material type. As the dependent variables, we recorded the participants' subjective ratings of the perceived intensity of the aforementioned five psychophysical dimensions under each type of haptic stimuli on each tested material. In each trial, a haptic stimulus (i.e., vibrotactile or pneumatic stimulus) lasted for 5 seconds followed by a 5-second resetting period for the skin to return to the neutral status. Including the time that the participant spent on ratings, the actual break lasted for around 20 seconds before the next trial. Noted that there were no visual and auditory cues during the stimuli for the participant. The Surface Pro screen in front of the participant showed the countdown process and presented the five sliders after each stimulus. The participant could move the slider to rate his/her perceived intensity on a continuous scale (1.00 - 100.00) for the five psychophysical haptic dimensions, with 1.00 indicating the lowest level and 100.00 as the highest level. 
The slider was initially placed in the middle (50.00 out of 100.00) of the scale which represents the sensation of the baseline material (i.e., plywood). Each participant performed the experiment in one sitting posture, including breaks. Each stimulus was repeated five times, and the first time was treated as a training session where no data was collected. Six testing materials (i.e., glass, paper, cotton, ceramic, wood and leather) were presented in a Latin-square-based counterbalanced order, and all the ViboPneumo signals were presented in a random order for each material. In total, each participant did a total of 6 materials $\times$ 7 stimuli $\times$ 5 repetitions = 210 trials. The total experimental time lasted no more than 2 hours.

\subsection{Task and Procedure}
We followed the common procedure of ``Introduction - PreQuestionnaire - Training - Testing" in the previous perception experiments \cite{asano2012vibrotactile, asano2013toward, asano2014toward, asano2014vibrotactile, cai2020thermairglove}. Our experiment involved one experimenter and one participant with two sessions: one training session and one experimental session. The training session was the same as the experimental session without data recording. The experiment started by introducing the process of the experiment, after which the participant was seated in a comfortable position and finished a pre-questionnaire about the demographic information. The participant needed to wash their hands with soap and dry them with a towel to ensure normal tactile perception on his/her bare fingerpad before the experiment. Then the experimenter measured the width of the DIP joint of the index finger on the participant's dominant hand and assigned a fitting size of 3D-printed housing to the participant for wearing ViboPneumo. 

During the experiment, the participant was required to gently move the fingertip from side to side and tried to maintain the sliding speed of about 50 to 200 mm/s and normal pressure of about 0.3 to 1.2 N \cite{romano2011creating} as much as possible for clear texture perception on each textured surface. Before testing, the participant could practice his/her sliding movement with the aforementioned speed and pressing force range as much as possible, until s/he self-reported that s/he was ready for the experiment. 
The baseline stimulus was presented to the participants after every six trials \cite{yoshimoto2015material}. Noted that the participant could feel the texture of the baseline material (plywood) anytime when s/he wanted. Between every two types of material, the participant was instructed to take off the ViboPneumo device for a 3-minute compulsory rest. 

\subsection{Participants}

We recruited twelve participants (five females and seven males) from a local university, all of whom were right-handed and had no prior experience with wearable haptic interfaces. The participants had an average age of 29.1 years old (SD = 3.03), and the average width of the distal interphalangeal (DIP) of their dominant index fingers was 14.9 mm (SD = 1.33). {We conducted a power analysis suggesting the estimated sample size of 12 with an 89.2\% chance of perceiving different levels of roughness among different haptic stimuli at the 0.05 significance level.}

\subsection{Results}

\begin{figure}[t]
 \centering 
 \includegraphics[width=\columnwidth]{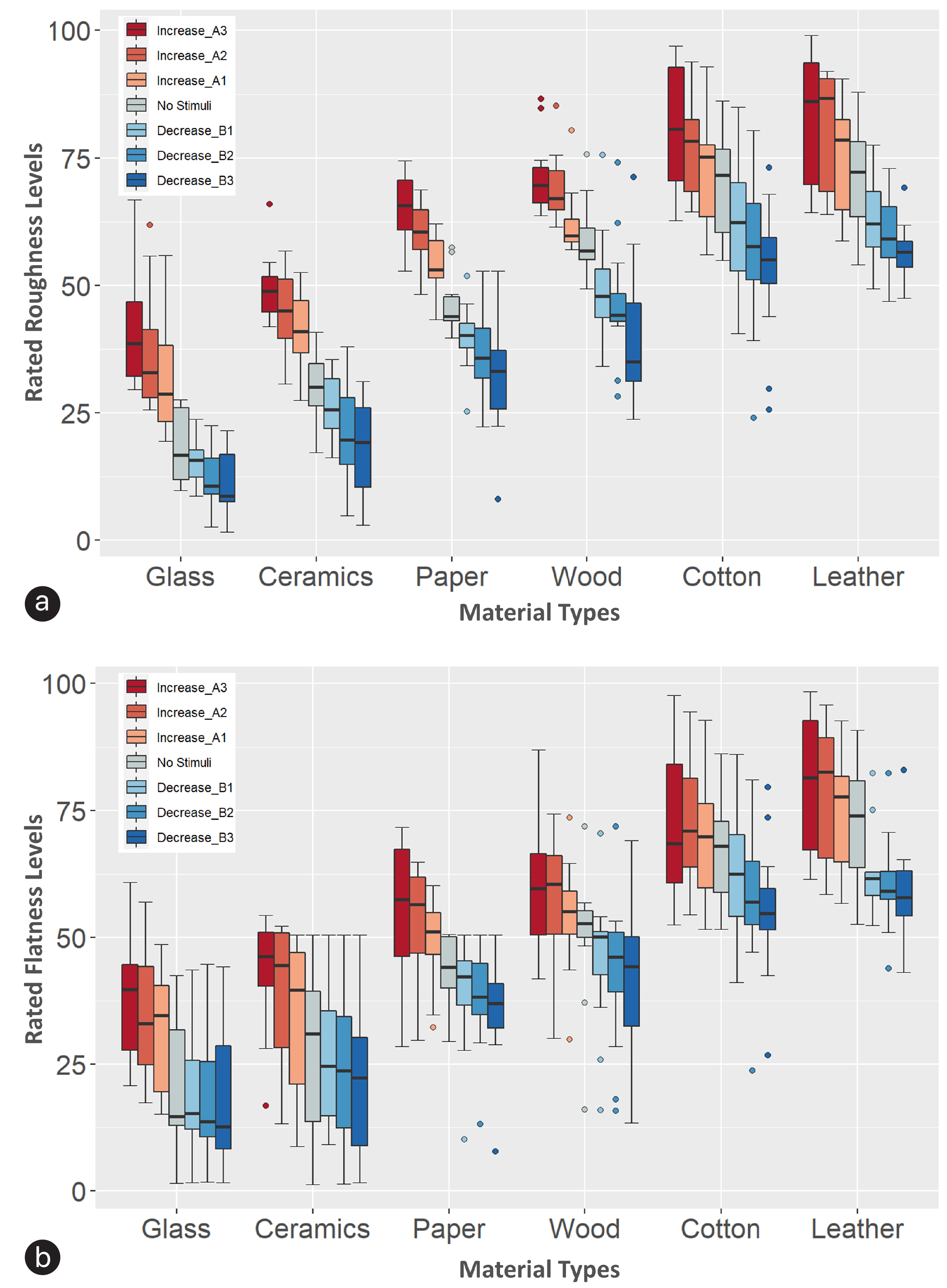}
 \caption{The results of participants' subjective rating levels of (a) perceived roughness and (b) perceived flatness. The legends include Increased\textunderscore A1, Increased\textunderscore A2, and Increased\textunderscore A3 representing the three amplitude levels of vibrotactile stimuli, and Decreased\textunderscore B1, Decreased\textunderscore B2, and Decreased\textunderscore B3 for the three pneumatic actuation levels, and No Stimuli for the no-stimuli condition.}
 \vspace{-0.1in}
 \label{result1}
 \end{figure}

\subsubsection{Subjective Ratings on Roughness}
Taking the stimuli and material types as independent factors, we ran a two-way repeated-measures ANOVA on the participants' ratings of perceived roughness among all the testing materials. Initially, we conducted Mauchly’s Sphericity Test and found a violation of the sphericity ($p <$ 0.05), so we applied the Greenhouse-Geisser correction to adjust the Degrees of Freedom. We found both the haptic stimuli ($F(1.47,15.475) = 79.502$, $p <$ 0.0001, $\eta_{p}^2$ = 0.878) and the material types ($F(2.225,24.477) = 64.279$, $p <$ 0.0001, $\eta_{p}^2$ = 0.854) had a statistically significant effect on the subjective rating values. 
Additionally, we found no significant interaction effect between these two factors. In terms of different ViboPneumo stimuli within each material, post-hoc pairwise comparison revealed significant differences in the user-rated values between almost all the pairs of stimuli and no stimuli conditions ($p <$ 0.05). In general, the material was perceived as rougher under the vibrotactile stimuli, with the roughness rating increasing with the amplitude level, and smoother under the pneumatic stimuli (i.e., the higher the air pressure, the smoother it was rated). Table \ref{pairwise} shows the detailed results of pairwise comparisons with Bonferroni correction for each condition and Table \ref{mean} shows the mean and SD values of participants’ subjective rating levels of perceived roughness. These findings suggested that the participants could perceive a range of roughness levels based on psychophysical dimensions, both increases and decreases, modulated by ViboPneumo under different stimuli.

\begin{table}[htbp]
\caption{Details of stimuli factor's effect on roughness and flatness per material. The ``$>$'' indicates the significant difference with p $<$ 0.05 {with Bonferroni correction}, and the ``$\sim$'' indicates no-significant difference.}
\resizebox{\columnwidth}{!}{\begin{tabular}{c|c|c|}
\cline{2-3}
\multirow{2}{*}{}              & Roughness                                                                                                                                                                   & Flatness                                                                                                                                                                                \\ \cline{2-3} 
                               & Stimuli                                                                                                                                                                     & Stimuli                                                                                                                                                                                 \\ \hline
\multicolumn{1}{|c|}{Glass}    & \begin{tabular}[c]{@{}c@{}}A2{$\sim$}A3\textgreater{}A1\textgreater{}N$\sim$B1$\sim$B2$\sim$B3,  \\ A2$\sim$A1, A2\textgreater{}N, A1\textgreater{}B1 \\B1$\sim$B3, N\textgreater{}B2/B3\end{tabular}     & \begin{tabular}[c]{@{}c@{}}A3\textgreater{}A2{$\sim$}A1\textgreater{}N$\sim$B1$\sim$B2$\sim$B3  \\ A2\textgreater{}N, A3\textgreater{}A1, N$\sim$B1/B2/B3, \\B1$\sim$B3, A1\textgreater{}B1/B2/B3 \end{tabular}                                         \\ \hline

\multicolumn{1}{|c|}{Ceramics} & \begin{tabular}[c]{@{}c@{}}A2{$\sim$}A3\textgreater{}A1\textgreater{}N{$\sim$}B1\textgreater{}B3{$\sim$}B2   \\ A2$\sim$A1, A2\textgreater{}N, A1\textgreater{}B1, \\B1\textgreater{}B2, N\textgreater{}B2/B3 \end{tabular}                                                                       & \begin{tabular}[c]{@{}c@{}}A3{$\sim$}A2\textgreater A1\textgreater{}N$\sim$B1{$\sim$}B2$\sim$B3 \\ A3\textgreater{}A1, A2\textgreater{}N, N$\sim$B1/B2/B3, \\ A1\textgreater{}B1/B2/B3, B1$\sim$B3\end{tabular}                     \\ \hline

\multicolumn{1}{|c|}{Paper}    & \begin{tabular}[c]{@{}c@{}} A3\textgreater{}A2\textgreater{}A1\textgreater{}N{$\sim$}B1{$\sim$}B2\textgreater{}B3   \\ A1\textgreater{}B1/B2, N\textgreater{}B2/B3, B1\textgreater{}B3          \end{tabular}                                                               & \begin{tabular}[c]{@{}c@{}}A3{$\sim$}A2$\sim$A1\textgreater{}N{$\sim$}B1$\sim$B2{$\sim$}B3 \\ A3{$\sim$}A1, A2\textgreater{}N, B1$\sim$B3, \\ A1\textgreater{}B1/B2/B3, N$\sim$B1/B2/B3\end{tabular}                     \\ \hline

\multicolumn{1}{|c|}{Wood}     & \begin{tabular}[c]{@{}c@{}}A3{$\sim$}A2\textgreater{}A1\textgreater{}N\textgreater{}B1$\sim$B2\textgreater{}B3, \\ A3\textgreater{}A1, A2\textgreater{}N, N\textgreater{}B2/B3, B1\textgreater{}B3\end{tabular} & \begin{tabular}[c]{@{}c@{}}A3{$\sim$}A2$\sim$A1\textgreater{}N$\sim$B1{$\sim$}B2$\sim$B3  \\ A3/A2\textgreater{}N, A3$\sim$A1, N$\sim$B2/B3, \\B1$\sim$B3, A1\textgreater{}B2/B3\end{tabular}                     \\ \hline

\multicolumn{1}{|c|}{Cotton}   & \begin{tabular}[c]{@{}c@{}} A3{$\sim$}A2\textgreater{}A1{$\sim$}N\textgreater{}B1\textgreater{}B2{$\sim$}B3  \\ A3\textgreater{}A1/N, A2\textgreater{}N, \\N\textgreater{}B2/B3, B2$\sim$B3 \end{tabular}                                                                         & \begin{tabular}[c]{@{}c@{}}A3{$\sim$}A2$\sim$A1$\sim$N\textgreater{}B2{$\sim$}B3\textgreater{}B1  \\ N\textgreater{}B2/B3, A3/A2\textgreater{}N, \\ A3$\sim$A1, B1$\sim$B2 \\ \end{tabular}                    \\ \hline

\multicolumn{1}{|c|}{Leather}  & \begin{tabular}[c]{@{}c@{}}A3{$\sim$}A2\textgreater{A1$\sim$}N\textgreater{}B1{$\sim$}B2\textgreater{}B3 \\ A3\textgreater{}A1/N, A2\textgreater{}N, \\N\textgreater{}B2/B3, B1\textgreater{}B3 \end{tabular} & \begin{tabular}[c]{@{}c@{}}A3{$\sim$}A2$\sim$A1{$\sim$}N\textgreater{}B1{$\sim$}B2$\sim$B3 \\ N\textgreater{}B1/B2/B3, A3$\sim$A1, \\ A2/A3\textgreater{}N,  B1$\sim$B3\end{tabular} \\ \hline
\end{tabular}
}
\label{pairwise}
 \vspace{-0.1in}
\end{table}

\begin{table}[h]
\caption{The mean and SD values of participants’ subjective rating levels of perceived roughness.} 
\resizebox{\columnwidth}{!}{\begin{tabular}{ccccccc}
\cline{2-7}
\multicolumn{1}{c|}{}               & \multicolumn{1}{c|}{Glass}                                                  & \multicolumn{1}{c|}{Ceramic}                                               & \multicolumn{1}{c|}{Paper}                                                 & \multicolumn{1}{c|}{Wood}                                                   & \multicolumn{1}{c|}{Cotton}                                                 & \multicolumn{1}{c|}{Leather}                                                \\ \hline
\multicolumn{1}{|c|}{Increase\_A3} & \multicolumn{1}{c|}{\begin{tabular}[c]{@{}c@{}}41.7\\ (12.57)\end{tabular}} & \multicolumn{1}{c|}{\begin{tabular}[c]{@{}c@{}}49.7\\ (6.54)\end{tabular}} & \multicolumn{1}{c|}{\begin{tabular}[c]{@{}c@{}}64.9\\ (7.38)\end{tabular}} & \multicolumn{1}{c|}{\begin{tabular}[c]{@{}c@{}}71.2\\ (7.54)\end{tabular}}  & \multicolumn{1}{c|}{\begin{tabular}[c]{@{}c@{}}82.2\\ (12.67)\end{tabular}} & \multicolumn{1}{c|}{\begin{tabular}[c]{@{}c@{}}84.1\\ (12.48)\end{tabular}} \\ \hline
\multicolumn{1}{|c|}{Increase\_A2} & \multicolumn{1}{c|}{\begin{tabular}[c]{@{}c@{}}36.8\\ (12.5)\end{tabular}}  & \multicolumn{1}{c|}{\begin{tabular}[c]{@{}c@{}}44.7\\ (8.43)\end{tabular}} & \multicolumn{1}{c|}{\begin{tabular}[c]{@{}c@{}}60.5\\ (6.03)\end{tabular}} & \multicolumn{1}{c|}{\begin{tabular}[c]{@{}c@{}}69.2\\ (6.71)\end{tabular}}  & \multicolumn{1}{c|}{\begin{tabular}[c]{@{}c@{}}77.7\\ (8.86)\end{tabular}}  & \multicolumn{1}{c|}{\begin{tabular}[c]{@{}c@{}}82.5\\ (11.05)\end{tabular}} \\ \hline
\multicolumn{1}{|c|}{Increase\_A1} & \multicolumn{1}{c|}{\begin{tabular}[c]{@{}c@{}}31.9\\ (11.72)\end{tabular}} & \multicolumn{1}{c|}{\begin{tabular}[c]{@{}c@{}}40.8\\ (8.34)\end{tabular}} & \multicolumn{1}{c|}{\begin{tabular}[c]{@{}c@{}}54.2\\ (5.45)\end{tabular}} & \multicolumn{1}{c|}{\begin{tabular}[c]{@{}c@{}}62.4\\ (6.54)\end{tabular}}  & \multicolumn{1}{c|}{\begin{tabular}[c]{@{}c@{}}73.3\\ (10.75)\end{tabular}} & \multicolumn{1}{c|}{\begin{tabular}[c]{@{}c@{}}77.5\\ (10.56)\end{tabular}} \\ \hline
\multicolumn{1}{|c|}{No Stimuli}   & \multicolumn{1}{c|}{\begin{tabular}[c]{@{}c@{}}17.9\\ (7.03)\end{tabular}}  & \multicolumn{1}{c|}{\begin{tabular}[c]{@{}c@{}}30.3\\ (6.89)\end{tabular}} & \multicolumn{1}{c|}{\begin{tabular}[c]{@{}c@{}}46.1\\ (5.78)\end{tabular}} & \multicolumn{1}{c|}{\begin{tabular}[c]{@{}c@{}}58.6\\ (7.10)\end{tabular}}  & \multicolumn{1}{c|}{\begin{tabular}[c]{@{}c@{}}70.8\\ (10.84)\end{tabular}} & \multicolumn{1}{c|}{\begin{tabular}[c]{@{}c@{}}74.1\\ (9.87)\end{tabular}}  \\ \hline
\multicolumn{1}{|c|}{Decrease\_B1} & \multicolumn{1}{c|}{\begin{tabular}[c]{@{}c@{}}15.7\\ (4.44)\end{tabular}}  & \multicolumn{1}{c|}{\begin{tabular}[c]{@{}c@{}}26.4\\ (6.19)\end{tabular}} & \multicolumn{1}{c|}{\begin{tabular}[c]{@{}c@{}}41.0\\ (4.72)\end{tabular}} & \multicolumn{1}{c|}{\begin{tabular}[c]{@{}c@{}}49.3\\ (10.83)\end{tabular}} & \multicolumn{1}{c|}{\begin{tabular}[c]{@{}c@{}}64.2\\ (14.66)\end{tabular}} & \multicolumn{1}{c|}{\begin{tabular}[c]{@{}c@{}}61.90\\ (7.08)\end{tabular}} \\ \hline
\multicolumn{1}{|c|}{Decrease\_B2} & \multicolumn{1}{c|}{\begin{tabular}[c]{@{}c@{}}13.0\\ (6.27)\end{tabular}}  & \multicolumn{1}{c|}{\begin{tabular}[c]{@{}c@{}}20.5\\ (9.61)\end{tabular}} & \multicolumn{1}{c|}{\begin{tabular}[c]{@{}c@{}}37.0\\ (7.16)\end{tabular}} & \multicolumn{1}{c|}{\begin{tabular}[c]{@{}c@{}}45.4\\ (13.16)\end{tabular}} & \multicolumn{1}{c|}{\begin{tabular}[c]{@{}c@{}}57.6\\ (15.34)\end{tabular}} & \multicolumn{1}{c|}{\begin{tabular}[c]{@{}c@{}}60.0\\ (7.47)\end{tabular}}  \\ \hline
\multicolumn{1}{|c|}{Decrease\_B3} & \multicolumn{1}{c|}{\begin{tabular}[c]{@{}c@{}}11.0\\ (6.71)\end{tabular}}  & \multicolumn{1}{c|}{\begin{tabular}[c]{@{}c@{}}17.9\\ (9.36)\end{tabular}} & \multicolumn{1}{c|}{\begin{tabular}[c]{@{}c@{}}31.9\\ (11.4)\end{tabular}} & \multicolumn{1}{c|}{\begin{tabular}[c]{@{}c@{}}38.6\\ (15.12)\end{tabular}} & \multicolumn{1}{c|}{\begin{tabular}[c]{@{}c@{}}53.4\\ (14.08)\end{tabular}} & \multicolumn{1}{c|}{\begin{tabular}[c]{@{}c@{}}58.0\\ (6.09)\end{tabular}}  \\ \hline
                                                   
\end{tabular}}
\label{mean}
\end{table}

Fig. \ref{result1}a depicts the descriptive results of participants' subjective ratings on the roughness perception, which varied across different stimuli and material types. We also observed certain levels of overlapping across different stimuli and materials. For instance, the box plot of Increase\_A1 for glass largely covers the range of the box plot for No Stimuli on ceramics, while the range of the box plot for paper with No Stimuli was overlapped with the box plot of wood with Decrease\_B1. This indicated that our device could provide a wide range of haptic stimuli that could potentially modify the perceived roughness of one type of material towards another type of material. For instance, with ViboPneumo, wood could be perceived as smooth as paper under the pneumatic stimuli which could reduce the wood's perceived roughness. By increasing the roughness with ViboPneumo, glass could be perceived as rough as ceramics.

\begin{figure}[htbp]
 \centering 
 \includegraphics[width=0.96\columnwidth]{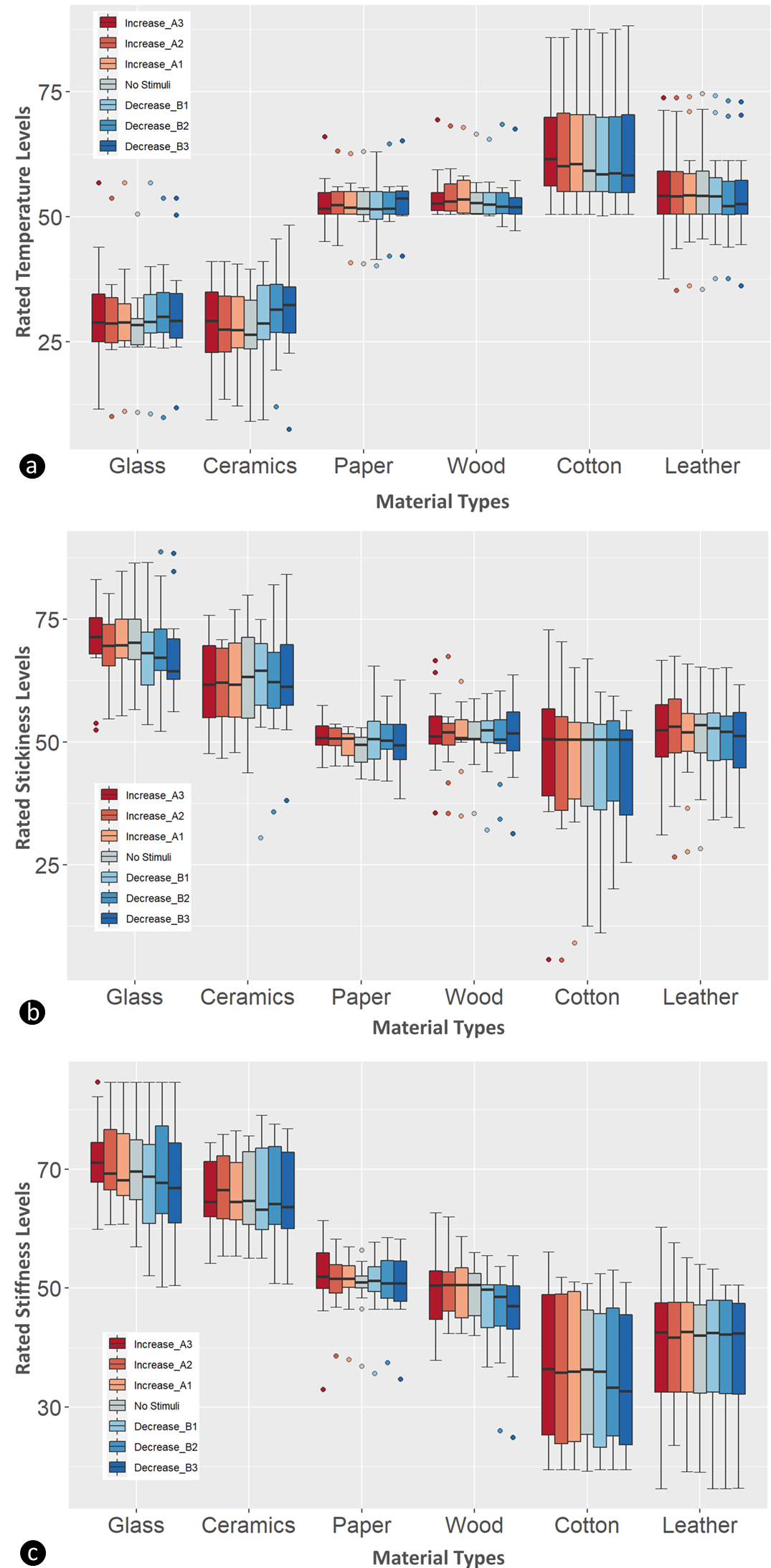}
 \caption{The results of participants' subjective rating levels of (a) perceived temperature (b) perceived stiffness and (c) perceived stickiness. The legends are same with Fig. \ref{result1}.}
 \vspace{-0.1in}
 \label{result1-2}
 \end{figure}

\subsubsection{Multi-Dimensional Haptic Properties Analysis}

Besides roughness, we were also interested in other haptic properties perceived by the participants, including flatness, temperature, stickiness, and stiffness. Fig. \ref{result1}b shows the descriptive results of subjective ratings on the perceived flatness, which demonstrates a similar trend compared to the results of roughness rating (Fig. \ref{result1}a). We ran a two-way repeated-measures ANOVA on the participants' flatness ratings, and the results showed that both the material type ($F(1.969,21.656)$ = 29.399, $p<$ 0.0001, $\eta_{p}^2$ = 0.728) and the stimuli ($F(1.231,13.538)$ = 29.160, $p<$ 0.0001, $\eta_{p}^2$ = 0.726) yielded significant effects on the user-rated flatness. Furthermore, we found a strong correlation between perceived roughness and flatness ratings ($r(420)=0.882$, $p<$0.001) based on Pearson correlation analysis.

We also found significant differences in temperature ($F(5,55)$ = 31.424, $p<$ 0.0001, $\eta_{p}^2$ = 0.741), stickiness ($F(1.596,17.558)$ = 10.811, $p<$ 0.0001, $\eta_{p}^2$ = 0.496) and stiffness ($F(1.955,21.510)$ = 31.119, $p <$ 0.0001, $\eta_{p}^2$ = 0.739) across the material types. However, our analysis did not reveal any significant differences in temperature ($F(6,66) = 1.304$, $p =$ 0.268, $\eta_{p}^2$ = 0.106), stiffness ($F(6,66) = 2.163$, $p =$ 0.058, $\eta_{p}^2$ = 0.164), and stickiness ($F(6,66) = 0.493$, $p =$ 0.811, $\eta_{p}^2$ = 0.043) across different ViboPneumo haptic stimuli, shown in Fig. \ref{result1-2}a, b, and c for the respective plots. 
This suggests that ViboPneumo did not explicitly alter or, in other words, could preserve these perceived haptic properties (i.e., temperature, stickiness, and stiffness) of the material surfaces.

\section{User-perception Experiment 2: MR Experience with ViboPneumo}

With the above experiment validating the effectiveness of roughness modulation through ViboPneumo, we further conducted the second user-perception experiment to investigate how ViboPneumo may affect the user experience in MR environments.

\subsection{Apparatus and Stimuli}

We created an MR scene using Unity3D (2019.3.4.39f). The application used HoloLens 2 with a hand-tracking function, a 5-megapixel camera for finger velocity estimation, and the ViboPneumo system, as shown in Fig. \ref{fig:teaser}a. The MR scene consisted of six physical objects: a wooden board, two cups (one glass and one ceramic), two table mats (one cotton and one leather), and one paper-made box (as shown in Tab.\ref{table2}). In MR, to create the visual-haptic matching, we developed virtual objects aligned with those physical objects in terms of size but made with different virtual materials (Fig. \ref{fig:teaser}b). Specifically, we rendered a virtual glass/ceramics cup on a physical ceramics/glass cup, a virtual leather/cotton table mat over a physical cotton/leather table mat, and a virtual wooden cube aligned with the paper-made cube, with a virtual paper displayed on the surface of the wooden table (applied scenes shown in Tab.\ref{table2}). There were two modes of object manipulation: 1) using the bare finger without ViboPneumo (denoted as BareFinger), and 2) wearing ViboPneumo with corresponding haptic feedback (denoted as ViboPneumo).

Based on the results of our first user-perception experiment, we determined the appropriate haptic stimuli for roughness modulation as follows: deploying vibrotactile stimuli to increase the perceived roughness on the glass cup for the virtual ceramic cup, the paper-made cube for the virtual wooden cube, and the cotton table mat for the virtual leather mat; and the pneumatic actuation to decrease the perceived roughness on the ceramic cup for the virtual glass cup, the wooden board for the virtual paper, and the leather table mat for the virtual cotton mat. According to our first experimental results and also considering flatness perception, we designed power-friendly parameters for the intensity of haptic stimuli, shown in Tab.\ref{table2}.

\subsection{Task and Procedure}


Each session included one participant and one experimenter. The experimenter introduced the instructions and procedures for the experiment and assisted participants in wearing the HoloLens 2 headset and the appropriate ViboPneumo devices. Participants were then instructed on how to use the ViboPneumo device to interact with objects in the MR environment. They experienced two testing sub-sessions of MR interaction that represented the two object-manipulation modes: BareFinger and ViboPneumo. In the ViboPneumo mode, participants were able to touch and feel the surface texture of various virtual objects through haptic feedback generated from ViboPneumo, which altered the physical objects. The order of the two modes was counterbalanced for each participant.

At the end of each sub-session, the participant completed a customized questionnaire derived from the presence questionnaire \cite{witmer1998measuring} and the system usability scale (SUS) \cite{bangor2008empirical}, using a 7-point Likert scale (1: strongly disagree - 7: strongly agree), {mainly focusing on visual-haptic matching experience in MR.} After the experiment session, we conducted a semi-structured interview with each participant to gather their qualitative feedback on ViboPneumo and its potential application. 

 \begin{table}[h]
\caption{{The applied MR scenes and parameters of haptic stimuli.}}
\resizebox{\columnwidth}{!}{
\begin{tabular}{c|c|c|c|}
\cline{2-4}
                                                                                                                              & Physical objects  & Virtual objects   & Applied stimuli \\ \hline
\multicolumn{1}{|c|}{\multirow{3}{*}{\begin{tabular}[c]{@{}c@{}}Roughness increasing\\ (Acceleration m/s$^2$)\end{tabular}}} & Glass cup         & Ceramics cup      & 3.7 m/s$^2$         \\ \cline{2-4} 
\multicolumn{1}{|c|}{}                                                                                                        & Paper-made box    & Wooden board      & 4.9 m/s$^2$         \\ \cline{2-4} 
\multicolumn{1}{|c|}{}                                                                                                        & Cotton table mat  & Leather table mat & 4.9 m/s$^2$         \\ \hline
\multicolumn{1}{|c|}{\multirow{3}{*}{\begin{tabular}[c]{@{}c@{}}Roughness decreasing\\ (Air pressure kpa)\end{tabular}}}  & Ceramics cup      & Glass cup         & 10 kpa          \\ \cline{2-4} 
\multicolumn{1}{|c|}{}                                                                                                        & Wooden board      & Paper-made box    & 8 kpa           \\ \cline{2-4} 
\multicolumn{1}{|c|}{}                                                                                                        & Leather table mat & Cotton table mat  & 6 kpa           \\ \hline
\end{tabular}
}
\vspace{-0.1in}
\label{table2}
\end{table}

\subsection{Participants}

We recruited ten participants (five females and five males) in this experiment, with an average age of 29.3 years (SD = 2.75). All the participants were right-handed and did not attend the previous experiments. One of them had previous MR experience but without any experience using haptic devices in MR. The average width of the DIP of the index fingers was 15.7 mm (SD = 1.73). {A power analysis indicated a sample size of 5 with a 96.1\% chance for perceiving the difference between two modes of object manipulation with a 0.05 significance level.}

\subsection{Questionnaire Results}
Fig. \ref{fig:mrexperience} illustrates the comparison between the two object-manipulation modes. Taking the operation mode as the independent factor and the questionnaire responses as the dependent variables, we analyzed the results using the Wilcoxon signed-rank test. The results showed that the type of the operating mode significantly affected the participant-rated consistency of visual and haptic information ($Z = -2.816, p = 0.005$), consistency of touch experience in MR and the real world ($Z = -2.816, p=0.005$), perceived naturalness of the texture ($Z = -2.814, p=0.005$), the users' preference ($Z = -2.388, p=0.017$). There was no significant difference between these two conditions for the participants' responses to the questionnaire items on the capability of actively surveying or searching the MR environment ($Z = -0.378, p=0.705$) and usability ($Z = -0.816, p=0.414$), indicating that the ViboPneumo system did not constrain the users' exploration in MR environments and was easy to use for novice users.

\subsection{Qualitative Feedback}

Regarding qualitative feedback, we first asked the participants how they could distinguish different objects under the same category (e.g., glass cups and ceramic cups). All the participants mentioned that they could visually distinguish different table mats in MR, but found it difficult to differentiate different cups due to the overlapping between virtual and physical objects. However, the usage of ViboPneumo provided different roughness sensations on these materials to enhance their experience in discriminating between ceramic and glass cups. P3 commented, \textit{``I can discriminate these two different cups through texture interaction with reasonable haptic experience, but I feel extreme mismatching between visual and haptic experience when I use my bare finger, which is really unpleasant.''} We asked the participants about their feeling about the virtual wooden box overlaying the paper box. All the participants agreed that the paper box with an augmented wooden appearance felt like a wooden texture with ViboPneumo. 

Next, we encouraged the participants to find some possible applications for ViboPneumo. P8, who had MR experience commented, \textit{``I think it could be a useful tool for designers. Some designers use AR to demonstrate different appearances of customized/personalized products, such as clothes, where this system can not only change the appearances but also provide the corresponding haptic feedback of the fabrics.''} Another participant (P5) commented that it could be used for game control. {For example, using ViboPneumo, a user can control the movements of a virtual role on a physical material surface through MR while experiencing modified perceived roughness to create a matching experience (e.g., the perceived roughness can be decreased when the virtual role walks on a virtual icy surface).}

\begin{figure}[t]
 \centering 
 \includegraphics[width=\linewidth]{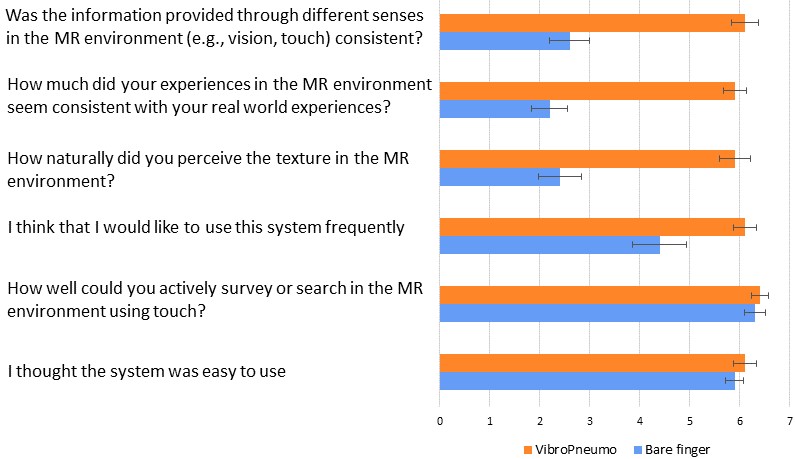}
 \vspace{-0.1in}
 \caption{Questionnaire responses on the MR experience.}
 \label{fig:mrexperience}
 \vspace{-0.1in}
 \end{figure}
 \vspace{-0.1in}

\section{Discussion}

Our first user-perception experiment examined the capability of altering perceived roughness through ViboPneumo among six selected materials: glass, ceramics, paper, wood, leather, and cotton. Our results showed that the vibrotactile stimuli generated by ViboPneumo could increase the perceived roughness of physical materials, with higher vibration amplitude resulting in a rougher perception, particularly for smoother materials such as glass and ceramics. These findings are consistent with previous research on altering roughness on physical objects \cite{asano2014vibrotactile, yoshimoto2015material}. In addition, ViboPneumo was able to reduce the contact area between the fingerpad skin and the material surface by lifting the fingerpad, which resulted in a decrease in perceived roughness. We found that a larger expanded displacement (i.e., higher air pressure levels) yielded a lower roughness perception. {We infer that two mechanisms may have contributed to this effect: 1) As we hypothesized, reducing contact area decreases the applied normal force on the textured surfaces \cite{van2015review}, enabling to decrease the perceived roughness \cite{lederman1972fingertip}; 2) the smaller skin area with fewer tactile mechanoreceptors yields overall lower levels of perceptual activity stimuli, and roughness perception is positively related to the activity levels of tactile mechanoreceptors \cite{blake1997neural}. We suspect that both of these two mechanisms jointly contributed to the perceived roughness reduction.} 


We also observed our device has influenced the perceived flatness (i.e., macro-roughness) of the materials. Okamoto et al. \cite{okamoto2012psychophysical} suggested that it was difficult for humans to distinguish these two dimensions/concepts: micro-roughness and macro-roughness (i.e., flatness) due to the significant overlap between the perceptual mechanisms and the mental models. During our tests, we also observed this phenomenon, with some participants confusing these two attributes. In our results, we found similar trends in the participant-perceived flatness and roughness, which echoes Gescheider et al.'s \cite{gescheider2005perception} finding that these two dimensions jointly influenced the perception of textured surfaces. Noted that all participants agreed that ViboPneumo yielded largely flatness decreasing for those bumpy materials, such as taurillon leather. One possible reason could be that the reduced contact area resulted in the decreasing of applied normal force on the leather surface, reducing the sensitivity of spatial patterns on the material surface related to macro roughness \cite{okamoto2012psychophysical, connor1992neural}.


Additionally, we did not observe a significant influence of ViboPneumo on the other three material properties (i.e., temperature, stickiness, and stiffness), suggesting ViboPneumo may potentially preserve these attributes during the physical surface-touching interaction. {Previous research has suggested that reducing the applied normal forces and contact area might decrease stickiness \cite{van2015review} or increase cold sensations \cite{jones2008warm} during contact with textured surfaces. However, our experiments did not show statistically significant changes. Regarding stickiness perception, we hypothesized that users might also feel the friction between the silicon tube and the material surface, which could mask any changes in perceived stickiness resulting from the reduced contact area. Furthermore, we instructed users to maintain a normal pressure within 0.3-1.2 N during finger sliding, resulting in a small actual contact area of the fingerpad skin on the materials. Such the small range of force and contact area variance might not be large enough to perceive significant changes in temperature as the previous studies \cite{greenspan1985primate, stevens1971spatial} have mainly focused on changes in perceived temperature with large contact areas (e.g., over 10 cm$^2$). Finally, we did not require participants to actively press on the textured surfaces to perceive stiffness, which may explain why we did not observe significant changes in perceived stiffness, as the applied forces were likely to be relatively steady, but the stiffness perception is related to the change rate of stress \cite{lawrence2000rate}.}


Our second experiment on MR experience revealed altering roughness through ViboPneumo could improve the visual-haptic matching experience, enhancing the object perception experience in MR. In a mixed-reality environment, AR allows virtual objects to exist in the real environment, and we expect the touching experience of the virtual object to be consistent with the visual appearance. Using the physical object as the haptic proxy for the virtual object in MR \cite{kwon2009effects, hochreiter2018cognitive, 10.1145/3313831.3376313}, it is important for the user to feel a matching visual-haptic experience. ViboPneumo can benefit from reusable haptic proxies for diverse haptic texture experiences in MR by altering the perceived roughness while preserving the other material attributes of the physical object. For example, a user can not only feel the original texture of a physical ceramic mug but also alter the roughness of the ceramic mug to feel like a glass or metal material through ViboPneumo in MR. Additionally, we collected some possible application cases for ViboPneumo from our participants. One straightforward application is for immersive gaming. Some participants also suggested applying ViboPneumo to enhance the museum experience. For example, a user can wear ViboPneumo and touch the glass shell surface with increased roughness through vibrotactile stimuli, to feel the roughness of the ancient pottery inside the glass showcase box (Fig. \ref{fig:teaser}e). Moreover, ViboPneumo can assist designers in evaluating a 3D-printed prototype and altering its roughness or switching the material of specific parts, without any extra fabrication requirements (Fig. \ref{fig:teaser}f). Moreover, ViboPneumo can also serve as a haptic design tool, aiding in customizing perceived roughness for physical materials. As Asano et al. claimed \cite{asano2014toward}, two industrial designers could discuss the texture design for a product cover (e.g., leather), and one of them could deliver customized preferences of material properties (e.g., rougher leather) to another designer through a haptic design tool (e.g., ViboPneumo), for perceiving the customized texture sensations based on the current product.





\section{Limitations and Future Work}

We identified a few limitations and potential for improvement in our current system. Firstly, our device required a customized size design to fit different users' fingers. We carefully designed the size of 3D-printed housing (i.e., width from 13 to 18 mm) to match the user's index finger. Based on our empirical adjustment, the distance between the bottom surface of the pneumatic actuator to the bottom of the user's fingerpad is about 4 mm, enabling effective lift of the fingerpad with a reduced contact area when air pressure exceeds 5 kPa. Currently, we adopted TPU (90A hardness) materials for housing fabrication, which allows clamping of the finger. We also attempted to use silicone for our housing fabrication, but the soft housing made the pneumatic actuator easy to wrinkle during inflation, causing unbalanced air pressure distribution inside the tube. Furthermore, it should be noted that our current device is designed to modulate roughness on physical props but may not be effective for physical surfaces smaller than the fingertip or very concave objects due to the shape-related limitation of the current haptic device. However, we believe that future research may explore ways to develop and assemble multiple mini pneumatic actuators to achieve the bendable finger-worn device to support more exploring space. Such an idea has been applied in soft robots to support sophisticated manipulation \cite{shintake2017soft, xavier2022soft}.

Our current system has limited support for roughness modulation on a small area of the user's skin, specifically the fingerpad, and it is challenging to expand it to cover larger areas like the palm. However, existing research \cite{kalantari2018exploring} has shown that the index finger and thumb are the most sensitive for haptic perception, suggesting the possibility of using multiple finger-worn haptic devices to support more manipulation gestures in MR environments. In the future, we plan to expand our exploration of roughness modulation to include more types of materials, such as resin or ABS material, as we showed in Fig. \ref{fig:teaser}(f), to support altering roughness of 3D-printed objects. In addition, we are also interested in exploring the activation of both vibratory and pneumatic actuation on ViboPneumo to achieve finer controlled roughness modulation on various physical materials, which could provide more diverse tactile sensations. Another interesting idea is to create a haptic database that includes ratings of multi-modal material properties from different materials and stimuli, as well as the overlapping among these modulated haptic stimuli in terms of roughness perception. Such a database could facilitate finding proper material replacements for visual-haptic exploration in MR environments. {Based on such the database, we can investigate the integration of more types of virtual materials with specific physical materials with roughness modulation, creating diverse visual-tactile combinations for the MR experience. In the future, we are also interested in exploring whether ViboPneumo can create the haptic perception of a new type of material through roughness modulation without visual cues.}

Another limitation of our study is the exclusion of the bare finger condition in the first experiment. This is because we were concerned that frequent wearing and removing the haptic device could potentially impact the user experience and fluency in the experiment. To mitigate any potential influence on texture perception, we meticulously adjusted the wearable haptic device for each participant, ensuring no contact or interference between the silicone tube and the material surface texture under the ``no stimuli'' condition. Additionally, the results of the second user-perception experiment demonstrated that wearing the ViboPneumo device resulted in an enhanced visual-haptic matching experience and a more natural texture perception in MR. This suggests that ViboPneumo effectively modulates roughness compared to the ``no stimuli'' condition. In the second user-perception experiment, we noted that there was the absence of a comparison with the vibration-only method. Note that simply adopting vibration on users' fingertips can only increase the roughness. However, our system goes beyond that by aiming to modulate the perceived roughness through the integration of both vibration and pneumatic actuation, allowing us to increase and decrease roughness as needed. As a result, we did not directly compare our approach with the vibration-only method in this study. In future work, it would be beneficial to include a direct comparison between our approach and the vibration-only mode to provide a comprehensive evaluation of the MR experience for commercial usage.


Lastly, we observed some variance among participants in their subjective ratings of perceived materials and stimuli during the first user-perception experiment. This variance could be the potential influence of gender or age group. Previous research has suggested that gender and age group may affect the perceptual levels of haptic stimuli \cite{lue2018thermal}. To investigate the potential influence of gender, we conducted a non-parametric one-way ANOVA test on the users' rating results for roughness and flatness, with gender as an independent factor. The Kruskal-Wallis H test showed that there was no significant difference in perceived roughness levels and perceived flatness levels between males and females. In addition, we notice that the age range of our user group is narrow, mainly consisting of young people (23-34). This might be because we suppose that young people reflect the major customer group of VR/MR. According to a statistical report, individuals aged 25-34 represent the demographic with the highest proportion (~35\%) of VR/MR users \cite{vrage}. However, we did not consider users in other age ranges, resulting in the limited age of user groups in effective usage with a small sample size in the experiment. In future studies, it would be interesting to investigate haptic perception with a larger sample size and diverse demographic information (e.g., ages or occupations) to further examine the potential influence of texture perception using ViboPneumo.







\section{Conclusion}

In this paper, we propose ViboPneumo, a finger-worn haptic device that utilizes vibrotactile and pneumatic feedback to alter the perceived roughness of physical objects in mixed reality. The device features a linear resonant actuator that increases roughness using vibrotactile stimuli and a hollow pneumatic actuator that decreases roughness by reducing the contact area between the user's fingerpad and the physical surface. Our user-perception experiments demonstrated that ViboPneumo can successfully alter the perceived roughness of certain materials. We also observed that the user-rated roughness levels overlapped across some materials, suggesting the potential to modulate the perceived roughness from one material to another. Our user studies on mixed reality experience showed that using ViboPneumo to alter the perceived roughness of physical materials significantly improved users' visual-haptic matching experience in MR. The ViboPneumo system can be applied to various MR scenarios that require visual-haptic exploration.

\section{Acknowledgments}
This research was partially supported by the Centre for Applied Computing and Interactive Media (ACIM) of School of Creative Media, City University of Hong Kong. This work was also partially supported by the National Natural Science Foundation of China (Project No. 62172346), the Guangdong Basic and Applied Basic Research Foundation (Project No. 2021A1515011893), the CityU Contract Research\_RMGS (Project No. 9239092), and the CityU Donations for Research Projects\_RMGS (Project No. 9229075), and the Innovation and Technology Fund (Project No. ITS/326/21FP).

\bibliographystyle{IEEEtran}
\bibliography{IEEEabrv, main}

\begin{thebibliography}{10}
\providecommand{\url}[1]{#1}
\csname url@samestyle\endcsname
\providecommand{\newblock}{\relax}
\providecommand{\bibinfo}[2]{#2}
\providecommand{\BIBentrySTDinterwordspacing}{\spaceskip=0pt\relax}
\providecommand{\BIBentryALTinterwordstretchfactor}{4}
\providecommand{\BIBentryALTinterwordspacing}{\spaceskip=\fontdimen2\font plus
\BIBentryALTinterwordstretchfactor\fontdimen3\font minus \fontdimen4\font\relax}
\providecommand{\BIBforeignlanguage}[2]{{%
\expandafter\ifx\csname l@#1\endcsname\relax
\typeout{** WARNING: IEEEtran.bst: No hyphenation pattern has been}%
\typeout{** loaded for the language `#1'. Using the pattern for}%
\typeout{** the default language instead.}%
\else
\language=\csname l@#1\endcsname
\fi
#2}}
\providecommand{\BIBdecl}{\relax}
\BIBdecl

\bibitem{chen2009exploring}
X.~Chen, F.~Shao, C.~Barnes, T.~Childs, and B.~Henson, ``Exploring relationships between touch perception and surface physical properties,'' \emph{International Journal of Design}, vol.~3, no.~2, pp. 67--76, 2009.

\bibitem{djonov2011semiotics}
E.~Djonov and T.~Van~Leeuwen, ``The semiotics of texture: From tactile to visual,'' \emph{Visual Communication}, vol.~10, no.~4, pp. 541--564, 2011.

\bibitem{yu2019skin}
X.~Yu, Z.~Xie, Y.~Yu, J.~Lee, A.~Vazquez-Guardado, H.~Luan, J.~Ruban, X.~Ning, A.~Akhtar, D.~Li \emph{et~al.}, ``Skin-integrated wireless haptic interfaces for virtual and augmented reality,'' \emph{Nature}, vol. 575, no. 7783, pp. 473--479, 2019.

\bibitem{bau2010teslatouch}
O.~Bau, I.~Poupyrev, A.~Israr, and C.~Harrison, ``Teslatouch: electrovibration for touch surfaces,'' in \emph{Proceedings of the 23nd annual ACM symposium on User interface software and technology}, 2010, pp. 283--292.

\bibitem{nakagaki2016materiable}
K.~Nakagaki, L.~Vink, J.~Counts, D.~Windham, D.~Leithinger, S.~Follmer, and H.~Ishii, ``Materiable: Rendering dynamic material properties in response to direct physical touch with shape changing interfaces,'' in \emph{Proceedings of the 2016 CHI Conference on Human Factors in Computing Systems}, 2016, pp. 2764--2772.

\bibitem{bochereau2018perceptual}
S.~Bochereau, S.~Sinclair, and V.~Hayward, ``Perceptual constancy in the reproduction of virtual tactile textures with surface displays,'' \emph{ACM Transactions on Applied Perception (TAP)}, vol.~15, no.~2, pp. 1--12, 2018.

\bibitem{whitmire2018haptic}
E.~Whitmire, H.~Benko, C.~Holz, E.~Ofek, and M.~Sinclair, ``Haptic revolver: Touch, shear, texture, and shape rendering on a reconfigurable virtual reality controller,'' in \emph{Proceedings of the 2018 CHI conference on human factors in computing systems}, 2018, pp. 1--12.

\bibitem{degraen2021capturing}
D.~Degraen, M.~Piovar{\v{c}}i, B.~Bickel, and A.~Kr{\"u}ger, ``Capturing tactile properties of real surfaces for haptic reproduction,'' in \emph{The 34th Annual ACM Symposium on User Interface Software and Technology}, 2021, pp. 954--971.

\bibitem{cai2020thermairglove}
S.~Cai, P.~Ke, T.~Narumi, and K.~Zhu, ``Thermairglove: A pneumatic glove for thermal perception and material identification in virtual reality,'' in \emph{2020 IEEE Conference on Virtual Reality and 3D User Interfaces (VR)}.\hskip 1em plus 0.5em minus 0.4em\relax IEEE, 2020, pp. 248--257.

\bibitem{zhu2022tapetouch}
L.~Zhu, X.~Jiang, J.~Shen, H.~Zhang, Y.~Mo, and A.~Song, ``Tapetouch: A handheld shape-changing device for haptic display of soft objects,'' \emph{IEEE Transactions on Visualization and Computer Graphics}, vol.~28, no.~11, pp. 3928--3938, 2022.

\bibitem{tiest2010tactual}
W.~M.~B. Tiest, ``Tactual perception of material properties,'' \emph{Vision research}, vol.~50, no.~24, pp. 2775--2782, 2010.

\bibitem{di2022roughness}
N.~Di~Stefano and C.~Spence, ``Roughness perception: A multisensory/crossmodal perspective,'' \emph{Attention, Perception, \& Psychophysics}, vol.~84, no.~7, pp. 2087--2114, 2022.

\bibitem{culbertson2014modeling}
H.~Culbertson, J.~Unwin, and K.~J. Kuchenbecker, ``Modeling and rendering realistic textures from unconstrained tool-surface interactions,'' \emph{IEEE transactions on haptics}, vol.~7, no.~3, pp. 381--393, 2014.

\bibitem{ando2007nail}
H.~Ando, E.~Kusachi, and J.~Watanabe, ``Nail-mounted tactile display for boundary/texture augmentation,'' in \emph{Proceedings of the international conference on Advances in computer entertainment technology}, 2007, pp. 292--293.

\bibitem{preechayasomboon2021haplets}
P.~Preechayasomboon and E.~Rombokas, ``Haplets: Finger-worn wireless and low-encumbrance vibrotactile haptic feedback for virtual and augmented reality,'' \emph{Frontiers in Virtual Reality}, vol.~2, p. 738613, 2021.

\bibitem{spagnoletti2018rendering}
G.~Spagnoletti, L.~Meli, T.~L. Baldi, G.~Gioioso, C.~Pacchierotti, and D.~Prattichizzo, ``Rendering of pressure and textures using wearable haptics in immersive vr environments,'' in \emph{2018 IEEE Conference on Virtual Reality and 3D User Interfaces (VR)}.\hskip 1em plus 0.5em minus 0.4em\relax IEEE, 2018, pp. 691--692.

\bibitem{kwon2009effects}
E.~Kwon, G.~J. Kim, and S.~Lee, ``Effects of sizes and shapes of props in tangible augmented reality,'' in \emph{2009 8th IEEE International Symposium on Mixed and Augmented Reality}.\hskip 1em plus 0.5em minus 0.4em\relax IEEE, 2009, pp. 201--202.

\bibitem{hochreiter2018cognitive}
J.~Hochreiter, S.~Daher, G.~Bruder, and G.~Welch, ``Cognitive and touch performance effects of mismatched 3d physical and visual perceptions,'' in \emph{2018 IEEE Conference on Virtual Reality and 3D User Interfaces (VR)}.\hskip 1em plus 0.5em minus 0.4em\relax IEEE, 2018, pp. 1--386.

\bibitem{10.1145/3313831.3376313}
\BIBentryALTinterwordspacing
Q.~Zhou, S.~Sykes, S.~Fels, and K.~Kin, ``Gripmarks: Using hand grips to transform in-hand objects into mixed reality input,'' in \emph{Proceedings of the 2020 CHI Conference on Human Factors in Computing Systems}, ser. CHI '20.\hskip 1em plus 0.5em minus 0.4em\relax New York, NY, USA: Association for Computing Machinery, 2020, p. 1–11. [Online]. Available: \url{https://doi.org/10.1145/3313831.3376313}
\BIBentrySTDinterwordspacing

\bibitem{10.1145/2858036.2858226}
\BIBentryALTinterwordspacing
M.~Azmandian, M.~Hancock, H.~Benko, E.~Ofek, and A.~D. Wilson, ``Haptic retargeting: Dynamic repurposing of passive haptics for enhanced virtual reality experiences,'' in \emph{Proceedings of the 2016 CHI Conference on Human Factors in Computing Systems}, ser. CHI '16.\hskip 1em plus 0.5em minus 0.4em\relax New York, NY, USA: Association for Computing Machinery, 2016, p. 1968–1979. [Online]. Available: \url{https://doi.org/10.1145/2858036.2858226}
\BIBentrySTDinterwordspacing

\bibitem{10.1145/3290605.3300923}
\BIBentryALTinterwordspacing
K.~Zhu, T.~Chen, F.~Han, and Y.-S. Wu, ``Haptwist: Creating interactive haptic proxies in virtual reality using low-cost twistable artefacts,'' in \emph{Proceedings of the 2019 CHI Conference on Human Factors in Computing Systems}, ser. CHI '19.\hskip 1em plus 0.5em minus 0.4em\relax New York, NY, USA: Association for Computing Machinery, 2019, p. 1–13. [Online]. Available: \url{https://doi.org/10.1145/3290605.3300923}
\BIBentrySTDinterwordspacing

\bibitem{asano2014toward}
S.~Asano, S.~Okamoto, Y.~Matsuura, and Y.~Yamada, ``Toward quality texture display: vibrotactile stimuli to modify material roughness sensations,'' \emph{Advanced Robotics}, vol.~28, no.~16, pp. 1079--1089, 2014.

\bibitem{pacchierotti2017wearable}
C.~Pacchierotti, S.~Sinclair, M.~Solazzi, A.~Frisoli, V.~Hayward, and D.~Prattichizzo, ``Wearable haptic systems for the fingertip and the hand: taxonomy, review, and perspectives,'' \emph{IEEE transactions on haptics}, vol.~10, no.~4, pp. 580--600, 2017.

\bibitem{bau2012revel}
O.~Bau and I.~Poupyrev, ``Revel: tactile feedback technology for augmented reality,'' \emph{ACM Transactions on Graphics (TOG)}, vol.~31, no.~4, pp. 1--11, 2012.

\bibitem{asano2013toward}
S.~Asano, S.~Okamoto, and Y.~Yamada, ``Toward augmented reality of textures: Vibrotactile high-frequency stimuli mask texture perception to be rougher or smoother?'' in \emph{2013 IEEE International Conference on Systems, Man, and Cybernetics}.\hskip 1em plus 0.5em minus 0.4em\relax IEEE, 2013, pp. 510--515.

\bibitem{yoshimoto2015material}
S.~Yoshimoto, Y.~Kuroda, M.~Imura, and O.~Oshiro, ``Material roughness modulation via electrotactile augmentation,'' \emph{IEEE Transactions on Haptics}, vol.~8, no.~2, pp. 199--208, 2015.

\bibitem{maeda2016wearable}
T.~Maeda, R.~Peiris, M.~Nakatani, Y.~Tanaka, and K.~Minamizawa, ``Wearable haptic augmentation system using skin vibration sensor,'' in \emph{Proceedings of the 2016 Virtual Reality International Conference}, 2016, pp. 1--4.

\bibitem{etzi2014textures}
R.~Etzi, C.~Spence, and A.~Gallace, ``Textures that we like to touch: An experimental study of aesthetic preferences for tactile stimuli,'' \emph{Consciousness and cognition}, vol.~29, pp. 178--188, 2014.

\bibitem{asano2014vibrotactile}
S.~Asano, S.~Okamoto, and Y.~Yamada, ``Vibrotactile stimulation to increase and decrease texture roughness,'' \emph{IEEE Transactions on Human-Machine Systems}, vol.~45, no.~3, pp. 393--398, 2014.

\bibitem{van2015review}
J.~van Kuilenburg, M.~A. Masen, and E.~van~der Heide, ``A review of fingerpad contact mechanics and friction and how this affects tactile perception,'' \emph{Proceedings of the Institution of Mechanical Engineers, Part J: Journal of engineering tribology}, vol. 229, no.~3, pp. 243--258, 2015.

\bibitem{lederman1972fingertip}
S.~J. Lederman and M.~M. Taylor, ``Fingertip force, surface geometry, and the perception of roughness by active touch,'' \emph{Perception \& psychophysics}, vol.~12, pp. 401--408, 1972.

\bibitem{maeda2022fingeret}
T.~Maeda, S.~Yoshida, T.~Murakami, K.~Matsuda, T.~Tanikawa, and H.~Sakai, ``Fingeret: A wearable fingerpad-free haptic device for mixed reality,'' in \emph{Proceedings of the 2022 ACM Symposium on Spatial User Interaction}, 2022, pp. 1--10.

\bibitem{bouzit2002rutgers}
M.~Bouzit, G.~Burdea, G.~Popescu, and R.~Boian, ``The rutgers master ii-new design force-feedback glove,'' \emph{IEEE/ASME Transactions on mechatronics}, vol.~7, no.~2, pp. 256--263, 2002.

\bibitem{vechev2019tactiles}
V.~Vechev, J.~Zarate, D.~Lindlbauer, R.~Hinchet, H.~Shea, and O.~Hilliges, ``Tactiles: Dual-mode low-power electromagnetic actuators for rendering continuous contact and spatial haptic patterns in vr,'' in \emph{2019 IEEE Conference on Virtual Reality and 3D User Interfaces (VR)}.\hskip 1em plus 0.5em minus 0.4em\relax IEEE, 2019, pp. 312--320.

\bibitem{withana2018tacttoo}
A.~Withana, D.~Groeger, and J.~Steimle, ``Tacttoo: A thin and feel-through tattoo for on-skin tactile output,'' in \emph{Proceedings of the 31st Annual ACM Symposium on User Interface Software and Technology}, 2018, pp. 365--378.

\bibitem{han2018hydroring}
T.~Han, F.~Anderson, P.~Irani, and T.~Grossman, ``Hydroring: Supporting mixed reality haptics using liquid flow,'' in \emph{Proceedings of the 31st Annual ACM Symposium on User Interface Software and Technology}, 2018, pp. 913--925.

\bibitem{mazursky2021magnetio}
A.~Mazursky, S.-Y. Teng, R.~Nith, and P.~Lopes, ``Magnetio: Passive yet interactive soft haptic patches anywhere,'' in \emph{Proceedings of the 2021 CHI Conference on Human Factors in Computing Systems}, 2021, pp. 1--15.

\bibitem{teng2021touch}
S.-Y. Teng, P.~Li, R.~Nith, J.~Fonseca, and P.~Lopes, ``Touch\&fold: A foldable haptic actuator for rendering touch in mixed reality,'' in \emph{Proceedings of the 2021 CHI Conference on Human Factors in Computing Systems}, 2021, pp. 1--14.

\bibitem{choi2012vibrotactile}
S.~Choi and K.~J. Kuchenbecker, ``Vibrotactile display: Perception, technology, and applications,'' \emph{Proceedings of the IEEE}, vol. 101, no.~9, pp. 2093--2104, 2012.

\bibitem{strohmeier2017generating}
P.~Strohmeier and K.~Hornb{\ae}k, ``Generating haptic textures with a vibrotactile actuator,'' in \emph{Proceedings of the 2017 CHI Conference on Human Factors in Computing Systems}, 2017, pp. 4994--5005.

\bibitem{kyung2008ubi}
K.-U. Kyung and J.-Y. Lee, ``Ubi-pen: a haptic interface with texture and vibrotactile display,'' \emph{IEEE Computer Graphics and Applications}, vol.~29, no.~1, pp. 56--64, 2008.

\bibitem{romano2011creating}
J.~M. Romano and K.~J. Kuchenbecker, ``Creating realistic virtual textures from contact acceleration data,'' \emph{IEEE Transactions on haptics}, vol.~5, no.~2, pp. 109--119, 2011.

\bibitem{schorr2017fingertip}
S.~B. Schorr and A.~M. Okamura, ``Fingertip tactile devices for virtual object manipulation and exploration,'' in \emph{Proceedings of the 2017 CHI conference on human factors in computing systems}, 2017, pp. 3115--3119.

\bibitem{chen2019fw}
D.~Chen, A.~Song, L.~Tian, L.~Fu, and H.~Zeng, ``Fw-touch: A finger wearable haptic interface with an mr foam actuator for displaying surface material properties on a touch screen,'' \emph{IEEE transactions on haptics}, vol.~12, no.~3, pp. 281--294, 2019.

\bibitem{yem2017wearable}
V.~Yem and H.~Kajimoto, ``Wearable tactile device using mechanical and electrical stimulation for fingertip interaction with virtual world,'' in \emph{2017 IEEE Virtual Reality (VR)}.\hskip 1em plus 0.5em minus 0.4em\relax IEEE, 2017, pp. 99--104.

\bibitem{jeon2009haptic}
S.~Jeon and S.~Choi, ``Haptic augmented reality: Taxonomy and an example of stiffness modulation,'' \emph{Presence}, vol.~18, no.~5, pp. 387--408, 2009.

\bibitem{hachisu2012augmentation}
T.~Hachisu, M.~Sato, S.~Fukushima, and H.~Kajimoto, ``Augmentation of material property by modulating vibration resulting from tapping,'' in \emph{International conference on human haptic sensing and touch enabled computer applications}.\hskip 1em plus 0.5em minus 0.4em\relax Springer, 2012, pp. 173--180.

\bibitem{tao2021altering}
Y.~Tao, S.-Y. Teng, and P.~Lopes, ``Altering perceived softness of real rigid objects by restricting fingerpad deformation,'' in \emph{The 34th Annual ACM Symposium on User Interface Software and Technology}, 2021, pp. 985--996.

\bibitem{holliins1993perceptual}
M.~Holliins, R.~Faldowski, S.~Rao, and F.~Young, ``Perceptual dimensions of tactile surface texture: A multidimensional scaling analysis,'' \emph{Perception \& psychophysics}, vol.~54, pp. 697--705, 1993.

\bibitem{hollins2000individual}
M.~Hollins, S.~Bensma{\"\i}a, K.~Karlof, and F.~Young, ``Individual differences in perceptual space for tactile textures: Evidence from multidimensional scaling,'' \emph{Perception \& Psychophysics}, vol.~62, pp. 1534--1544, 2000.

\bibitem{tiest2006analysis}
W.~M.~B. Tiest and A.~M. Kappers, ``Analysis of haptic perception of materials by multidimensional scaling and physical measurements of roughness and compressibility,'' \emph{Acta psychologica}, vol. 121, no.~1, pp. 1--20, 2006.

\bibitem{asano2012vibrotactile}
S.~Asano, S.~Okamoto, Y.~Matsuura, H.~Nagano, and Y.~Yamada, ``Vibrotactile display approach that modifies roughness sensations of real textures,'' in \emph{2012 IEEE RO-MAN: The 21st IEEE International Symposium on Robot and Human Interactive Communication}.\hskip 1em plus 0.5em minus 0.4em\relax IEEE, 2012, pp. 1001--1006.

\bibitem{ochiai2014diminished}
Y.~Ochiai, T.~Hoshi, J.~Rekimoto, and M.~Takasaki, ``Diminished haptics: Towards digital transformation of real world textures,'' in \emph{International Conference on Human Haptic Sensing and Touch Enabled Computer Applications}.\hskip 1em plus 0.5em minus 0.4em\relax Springer, 2014, pp. 409--417.

\bibitem{nittala2019like}
A.~S. Nittala, K.~Kruttwig, J.~Lee, R.~Bennewitz, E.~Arzt, and J.~Steimle, ``Like a second skin: Understanding how epidermal devices affect human tactile perception,'' in \emph{Proceedings of the 2019 CHI Conference on Human Factors in Computing Systems}, 2019, pp. 1--16.

\bibitem{shimoga1993survey}
K.~B. Shimoga, ``A survey of perceptual feedback issues in dexterous telemanipulation. ii. finger touch feedback,'' in \emph{Proceedings of IEEE Virtual Reality Annual International Symposium}.\hskip 1em plus 0.5em minus 0.4em\relax IEEE, 1993, pp. 271--279.

\bibitem{tanaka2023full}
Y.~Tanaka, A.~Shen, A.~Kong, and P.~Lopes, ``Full-hand electro-tactile feedback without obstructing palmar side of hand,'' in \emph{Proceedings of the 2023 CHI Conference on Human Factors in Computing Systems}, 2023, pp. 1--15.

\bibitem{okamura1998vibration}
A.~M. Okamura, J.~T. Dennerlein, and R.~D. Howe, ``Vibration feedback models for virtual environments,'' in \emph{Proceedings. 1998 IEEE International Conference on Robotics and Automation (Cat. No. 98CH36146)}, vol.~1.\hskip 1em plus 0.5em minus 0.4em\relax IEEE, 1998, pp. 674--679.

\bibitem{ujitoko2019modulating}
Y.~Ujitoko, Y.~Ban, and K.~Hirota, ``Modulating fine roughness perception of vibrotactile textured surface using pseudo-haptic effect,'' \emph{IEEE Transactions on Visualization and Computer Graphics}, vol.~25, no.~5, pp. 1981--1990, 2019.

\bibitem{brisben1999detection}
A.~Brisben, S.~Hsiao, and K.~Johnson, ``Detection of vibration transmitted through an object grasped in the hand,'' \emph{Journal of neurophysiology}, vol.~81, no.~4, pp. 1548--1558, 1999.

\bibitem{landin2010dimensional}
N.~Landin, J.~M. Romano, W.~McMahan, and K.~J. Kuchenbecker, ``Dimensional reduction of high-frequency accelerations for haptic rendering,'' in \emph{Haptics: Generating and Perceiving Tangible Sensations: International Conference, EuroHaptics 2010, Amsterdam, July 8-10, 2010. Proceedings}.\hskip 1em plus 0.5em minus 0.4em\relax Springer, 2010, pp. 79--86.

\bibitem{bradski2000opencv}
G.~Bradski, ``The opencv library.'' \emph{Dr. Dobb's Journal: Software Tools for the Professional Programmer}, vol.~25, no.~11, pp. 120--123, 2000.

\bibitem{okamoto2012psychophysical}
S.~Okamoto, H.~Nagano, and Y.~Yamada, ``Psychophysical dimensions of tactile perception of textures,'' \emph{IEEE Transactions on Haptics}, vol.~6, no.~1, pp. 81--93, 2012.

\bibitem{hollins2000evidence}
M.~Hollins and S.~R. Risner, ``Evidence for the duplex theory of tactile texture perception,'' \emph{Perception \& psychophysics}, vol.~62, no.~4, pp. 695--705, 2000.

\bibitem{witmer1998measuring}
B.~G. Witmer and M.~J. Singer, ``Measuring presence in virtual environments: A presence questionnaire,'' \emph{Presence}, vol.~7, no.~3, pp. 225--240, 1998.

\bibitem{bangor2008empirical}
A.~Bangor, P.~T. Kortum, and J.~T. Miller, ``An empirical evaluation of the system usability scale,'' \emph{Intl. Journal of Human--Computer Interaction}, vol.~24, no.~6, pp. 574--594, 2008.

\bibitem{blake1997neural}
D.~T. Blake, S.~S. Hsiao, and K.~O. Johnson, ``Neural coding mechanisms in tactile pattern recognition: the relative contributions of slowly and rapidly adapting mechanoreceptors to perceived roughness,'' \emph{Journal of Neuroscience}, vol.~17, no.~19, pp. 7480--7489, 1997.

\bibitem{gescheider2005perception}
G.~A. Gescheider, S.~J. Bolanowski, T.~C. Greenfield, and K.~E. Brunette, ``Perception of the tactile texture of raised-dot patterns: A multidimensional analysis,'' \emph{Somatosensory \& motor research}, vol.~22, no.~3, pp. 127--140, 2005.

\bibitem{connor1992neural}
C.~E. Connor and K.~O. Johnson, ``Neural coding of tactile texture: comparison of spatial and temporal mechanisms for roughness perception,'' \emph{Journal of Neuroscience}, vol.~12, no.~9, pp. 3414--3426, 1992.

\bibitem{jones2008warm}
L.~A. Jones and H.-N. Ho, ``Warm or cool, large or small? the challenge of thermal displays,'' \emph{IEEE Transactions on Haptics}, vol.~1, no.~1, pp. 53--70, 2008.

\bibitem{greenspan1985primate}
J.~D. Greenspan and D.~R. Kenshalo, ``The primate as a model for the human temperature-sensing system: 2. area of skin receiving thermal stimulation (spatial summation),'' \emph{Somatosensory research}, vol.~2, no.~4, pp. 315--324, 1985.

\bibitem{stevens1971spatial}
J.~C. Stevens and L.~E. Marks, ``Spatial summation and the dynamics of warmth sensation,'' \emph{Perception \& Psychophysics}, vol.~9, no.~5, pp. 391--398, 1971.

\bibitem{lawrence2000rate}
D.~A. Lawrence, L.~Y. Pao, A.~M. Dougherty, M.~A. Salada, and Y.~Pavlou, ``Rate-hardness: A new performance metric for haptic interfaces,'' \emph{IEEE Transactions on Robotics and Automation}, vol.~16, no.~4, pp. 357--371, 2000.

\bibitem{shintake2017soft}
J.~Shintake, H.~Sonar, E.~Piskarev, J.~Paik, and D.~Floreano, ``Soft pneumatic gelatin actuator for edible robotics,'' in \emph{2017 IEEE/RSJ International Conference on Intelligent Robots and Systems (IROS)}.\hskip 1em plus 0.5em minus 0.4em\relax IEEE, 2017, pp. 6221--6226.

\bibitem{xavier2022soft}
M.~S. Xavier, C.~D. Tawk, A.~Zolfagharian, J.~Pinskier, D.~Howard, T.~Young, J.~Lai, S.~M. Harrison, Y.~K. Yong, M.~Bodaghi \emph{et~al.}, ``Soft pneumatic actuators: A review of design, fabrication, modeling, sensing, control and applications,'' \emph{IEEE Access}, 2022.

\bibitem{kalantari2018exploring}
F.~Kalantari, D.~Gueorguiev, E.~Lank, N.~Bremard, and L.~Grisoni, ``Exploring fingers’ limitation of texture density perception on ultrasonic haptic displays,'' in \emph{Haptics: Science, Technology, and Applications: 11th International Conference, EuroHaptics 2018, Pisa, Italy, June 13-16, 2018, Proceedings, Part I 11}.\hskip 1em plus 0.5em minus 0.4em\relax Springer, 2018, pp. 354--365.

\bibitem{lue2018thermal}
Y.-J. Lue, H.-H. Wang, K.-I. Cheng, C.-H. Chen, and Y.-M. Lu, ``Thermal pain tolerance and pain rating in normal subjects: Gender and age effects,'' \emph{European Journal of Pain}, vol.~22, no.~6, pp. 1035--1042, 2018.

\bibitem{vrage}
\BIBentryALTinterwordspacing
I.~Blagojevic, ``Virtual reality statistics,'' 2023. [Online]. Available: \url{https://99firms.com/blog/virtual-reality-statistics/#gref}
\BIBentrySTDinterwordspacing

\end{thebibliography}


\begin{IEEEbiography}[{\includegraphics[width=1in,height=1.25in,clip,keepaspectratio]{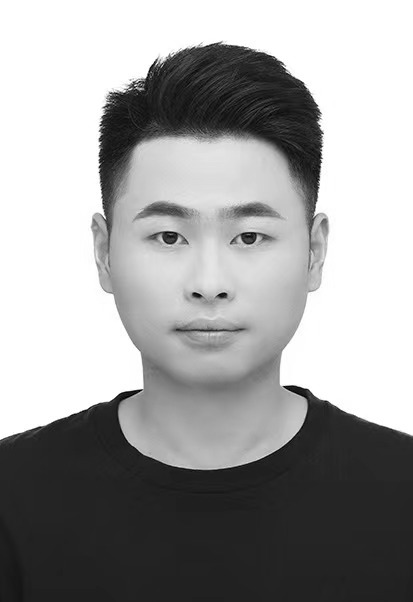}}]{Shaoyu Cai} is a Lecturer at Engineering Design \& Innovation Centre, College of Design and Engineering, National University of Singapore. He received his Ph.D. degree from the School of Creative Media, City University of Hong Kong, supervised by Prof. Kening Zhu. His research interest lies in the field of Human-Computer Interaction (HCI), with emphasis on cross-modal generation (e.g., from vision to touch), data-driven texture modeling and rendering, and haptic interface for VR/AR/MR.
\end{IEEEbiography}


\begin{IEEEbiography}[{\includegraphics[width=1in,height=1.25in,clip,keepaspectratio]{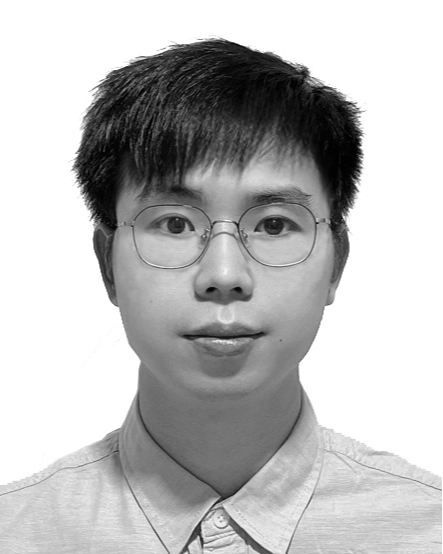}}]{Zhenlin Chen} received the B.E. degree from the Harbin Engineering University in 2017, and the M.S. degree from Harbin Institute of Technology in 2020. He is currently working toward the Ph.D. degree in biomedical engineering with the Department of Biomedical Engineering, City University of Hong Kong, Hong Kong. His research interests include microrobot manipulation system, human-robot Interaction.

\end{IEEEbiography}

\begin{IEEEbiography}[{\includegraphics[width=1in,height=1.25in,clip,keepaspectratio]{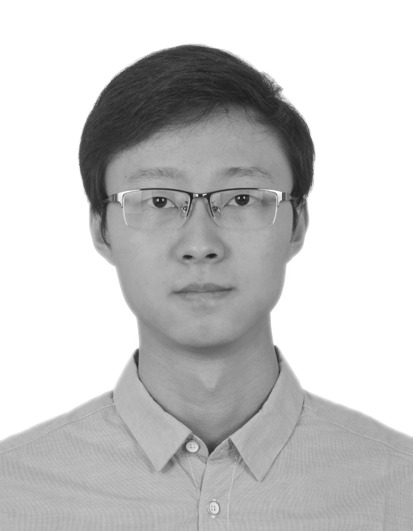}}]{Haichen Gao}
 is currently a PhD student in the School of Creative Media, City University of Hong Kong. His research interests are in the area of novel forms of interaction and haptic interface in virtual reality.
\end{IEEEbiography}

\begin{IEEEbiography}[{\includegraphics[width=1in,height=1.25in,clip,keepaspectratio]{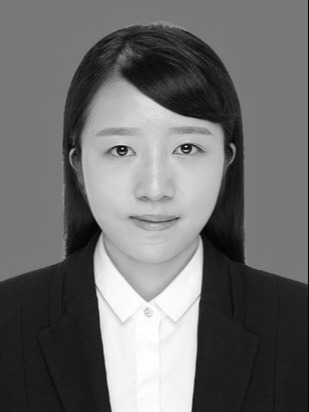}}]{Ya Huang} got her B.S. degree in Tsinghua University in 2015, and PhD degree in Tsinghua University in 2020. Now she is a Postdoctoral Fellow in the Department of Biomedical Engineering, City University of Hong Kong. Her current research interests mainly focus on wearable haptic feedback system for VR/AR and miniaturized electrostimulation therapy devices. 
\end{IEEEbiography}

\begin{IEEEbiography}[{\includegraphics[width=1in,height=1.25in,clip,keepaspectratio]{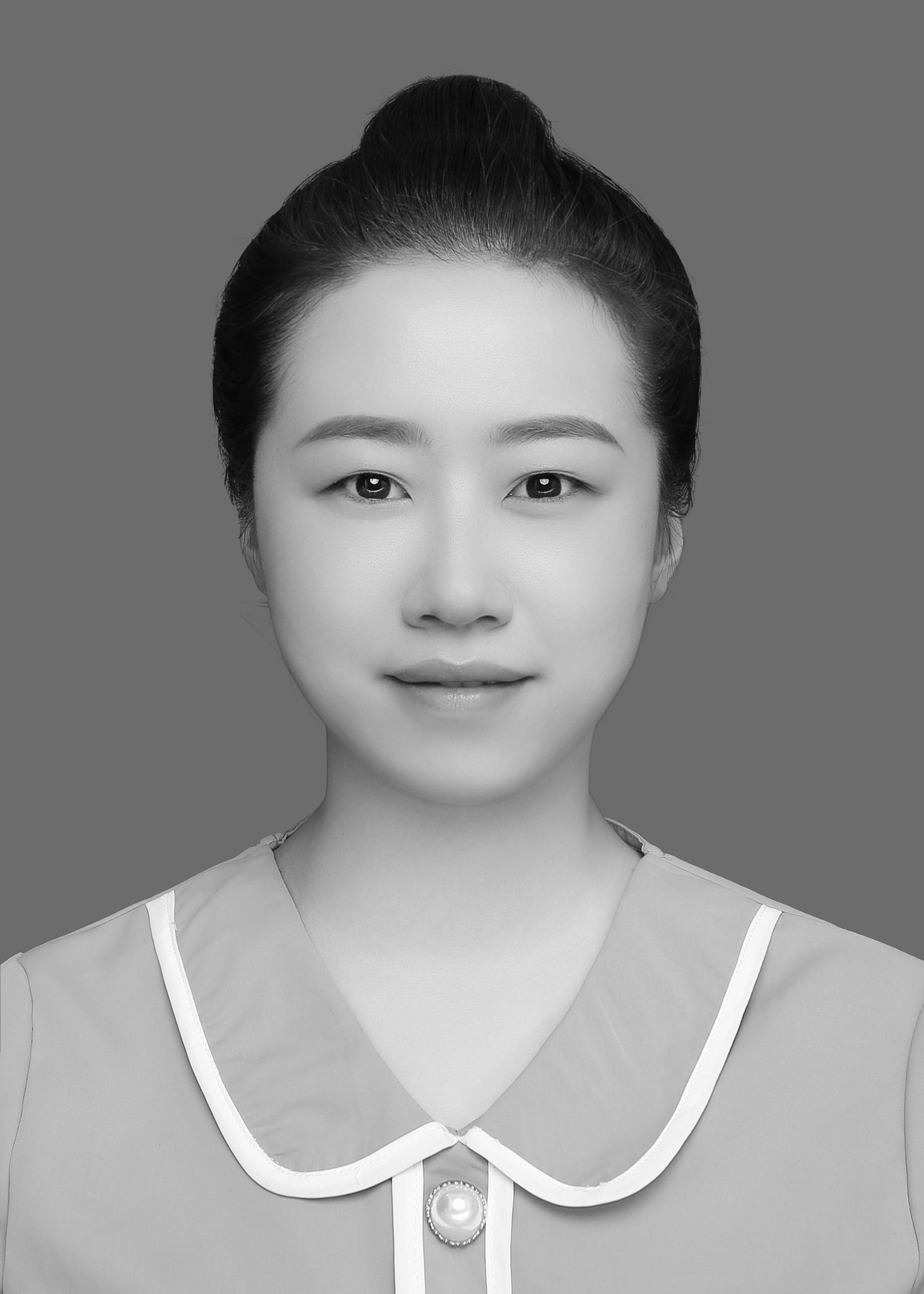}}]{Qi Zhang} is a research assistant in the School of Creative Media, City University of Hong Kong. Her research pursues the expedition of various human modalities by sensing technology, which involves creative materials employed in the novel interactive design space.
\end{IEEEbiography}

\begin{IEEEbiography}[{\includegraphics[width=1in,height=1.25in,clip,keepaspectratio]{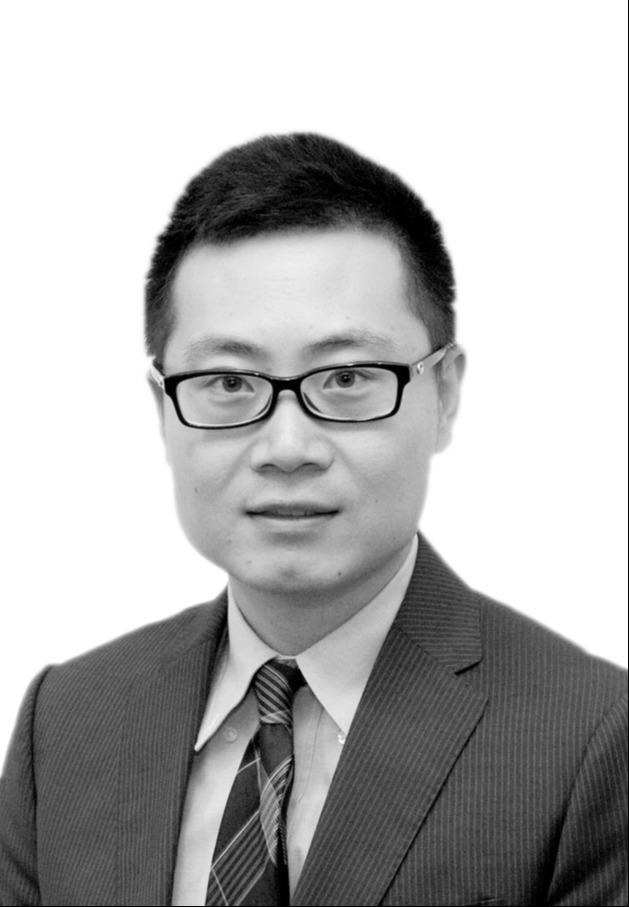}}]{Xinge Yu} is currently an Associate Professor of Biomedical Engineering at City University of Hong Kong (CityU), and Associate Director of Hong Kong Centre for Cerebro-cardiovascular Health Engineering. He is the recipient of Hong Kong RGC Fellow, NSFC Excellent Young Scientist Grant (Hong Kong \& Macao), Innovators under 35 China (MIT Technology Review), New Innovator of IEEE NanoMed, MINE Young Scientist Award, Gold Medal in the Inventions Geneva, CityU Outstanding Research Award, Stanford's top 2\% most highly cited scientists 2022 etc. Xinge Yu’s research group is focusing on skin-integrated electronics and systems for VR and biomedical applications. Now he serves the Associate Editor of Microsystem \& NanoEngineering and IEEE Open Journal of Nanotechnology; Editorial Boards for 14 journals, such as Soft Science, Materials Today Physics etc. He has published 160 papers in Nature, Nature Materials, Nature Biomedical Engineering, Nature Machine Intelligence, Nature Communications, Science Advances etc..
\end{IEEEbiography}

\begin{IEEEbiography}[{\includegraphics[width=1in,height=1.25in,clip,keepaspectratio]{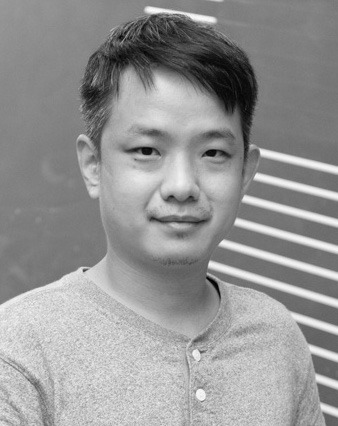}}]{Kening Zhu}
 is currently an associate professor jointly appointed by the School of Creative Media and the Department of Computer Science, City University of Hong Kong. He is leading the Multimodal and Embodied Interaction (MEI) Laboratory. He received his PhD from National University of Singapore. His research interests include tangible user interfaces, wearable user interfaces, mobile user interfaces, virtual and augmented reality, and the application of these interfaces/technologies in education, entertainment, accessibility, and so on.
\end{IEEEbiography}

\vfill


\end{document}